\documentclass[aps,prc,notitlepage,nofootinbib,amsmath,amssymb,showpacs,superscriptaddress,twocolumn]{revtex4-1}

\usepackage{graphics}
\usepackage{graphicx}
\usepackage{color}
\usepackage{hyperref}
\usepackage{bm}
\usepackage{calligra}
\usepackage[T1]{fontenc}
\usepackage{amsmath}
\usepackage{amssymb}
\usepackage{amsfonts}
\usepackage{mathrsfs}  
\usepackage{dsfont}
\usepackage{footmisc}

\usepackage[utf8]{inputenc}

\usepackage{lipsum}
\usepackage{textcomp}
\usepackage{color}

\newcommand{\trento}{T\raisebox{-.5ex}{R}ENTo }
\newcommand{\trentonosp}{T\raisebox{-.5ex}{R}ENTo}

\begin{document}

\title{Probing the structure of the initial state of heavy-ion collisions with $p_T$-dependent flow fluctuations}

\author{M.~Hippert}
\email{hippert@ifi.unicamp.br}
\affiliation{Instituto de F\'isica Gleb Wataghin, Universidade Estadual de Campinas, Rua S\'ergio Buarque de Holanda 777, 13083-859 S\~ao Paulo, Brazil}

\author{J.G.P.~Barbon}
\email{barbon@ifi.unicamp.br}
\affiliation{Instituto de F\'isica Gleb Wataghin, Universidade Estadual de Campinas, Rua S\'ergio Buarque de Holanda 777, 13083-859 S\~ao Paulo, Brazil}

\author{D.D.~Chinellato}
\email{daviddc@g.unicamp.br}
\affiliation{Instituto de F\'isica Gleb Wataghin, Universidade Estadual de Campinas, Rua S\'ergio Buarque de Holanda 777, 13083-859 S\~ao Paulo, Brazil}

\author{M.~Luzum}
\email{mluzum@usp.br}
\affiliation{Instituto de F\'{\i}sica, Universidade de  S\~{a}o Paulo,  Rua  do  Mat\~{a}o, 1371,  Butant\~{a},  05508-090,  S\~{a}o  Paulo,  Brazil}

\author{J.~Noronha}
\email{jn0508@illinois.edu}
\affiliation{Illinois Center for Advanced Studies of the Universe,\\ Department of Physics, 
University of Illinois at Urbana-Champaign, 1110 W. Green St., Urbana IL 61801-3080, USA}

\author{T.~Nunes da Silva}
\email{t.j.nunes@ufsc.br}
\affiliation{Departamento de  F\'{\i}sica - Centro de Ci\^encias  F\'{\i}sicas e Matem\'aticas, Universidade Federal de Santa Catarina, Campus Universit\'ario Reitor Jo\~ao David Ferreira Lima, Florian\'opolis 88040-900, Brazil}

\author{W.M.~Serenone}
\email{serenone@ifi.unicamp.br}
\affiliation{Instituto de F\'isica Gleb Wataghin, Universidade Estadual de Campinas, Rua S\'ergio Buarque de Holanda 777, 13083-859 S\~ao Paulo, Brazil}

\author{J.~Takahashi}
\email{jun@ifi.unicamp.br}
\affiliation{Instituto de F\'isica Gleb Wataghin, Universidade Estadual de Campinas, Rua S\'ergio Buarque de Holanda 777, 13083-859 S\~ao Paulo, Brazil}

\collaboration{{The ExTrEMe Collaboration}\textsuperscript{\hyperlink{extrm}{\S\S}}}

\date{\today}

\begin{abstract} 
The connection between initial-state geometry and anisotropic flow can be quantified through a well-established mapping between $p_T$-integrated flow harmonics and cumulants of the initial transverse energy distribution. In this paper we successfully extend this mapping to also include $p_T$-differential flow. In doing so, we find that subleading principal components of anisotropic flow can reveal previously unobserved details of the hydrodynamic response, in both the linear and the nonlinear regimes. Most importantly, we show that they provide novel information on the small-scale 
structures present in the initial stage of relativistic heavy-ion collisions. 
\end{abstract}

\maketitle

\begingroup
\setlength\footnotemargin{5pt}
\renewcommand{\footnotelayout}{\hspace{3.5pt}}
\renewcommand*{\thefootnote}{\fnsymbol{footnote}}
\footnotetext[10]{\hypertarget{extrm}{\href{https://sites.ifi.unicamp.br/extreme/}{The ExTrEMe  
Collaboration}} (\emph{``Experiment and Theory in Extreme MattEr''}) 
is a group of researchers focused on  phenomenology of High Energy Heavy Ion Collisions, with special interest in connecting theory with experiments. }
\renewcommand*{\thefootnote}{\arabic{footnote}}
\setcounter{footnote}{0}
\endgroup


\section{Introduction}

In high-energy heavy-ion collisions, the hydrodynamic expansion of the quark-gluon plasma (QGP) is driven by large pressure gradients 
that convert the anisotropic initial-state geometry into final-state momentum anisotropies, 
or anisotropic flow \cite{Ollitrault:1992bk}. 
In fact, a quantitative, event-by-event mapping between features of the initial geometry and 
the resulting anisotropic flow can be established in  hydrodynamic models of 
heavy-ion collisions \cite{Teaney:2010vd,Gardim:2011xv,Teaney:2012ke,Gardim:2014tya,Fu:2015wba,Rao:2019vgy}. 
Within this framework, it is possible to estimate how --- and to what extent --- 
anisotropic flow observables respond to initial-state fluctuations at 
different scales \cite{Gardim:2017ruc,Kozlov:2014fqa,Noronha-Hostler:2015coa}. 
The purpose of the present paper is to investigate the connection between 
subleading modes of anisotropic flow fluctuations  \cite{Bhalerao:2014mua,Mazeliauskas:2015vea,Mazeliauskas:2015efa,Cirkovic:2016kxt,Bozek:2017thv,Gardim:2019iah,Liu:2019jxg,Hippert:2019swu,Liu:2020ely,Sirunyan:2017gyb} 
and the aspects of the initial state of heavy-ion collisions, especially at smaller scales. 

The azimuthal flow can be characterized by flow harmonics $V_n$, which are defined as 
the Fourier coefficients of the azimuthal distribution of particles in a given event:
\begin{equation}
 \dfrac{d N}{dy \,p_T dp_T\, d\varphi} =\dfrac{1}{2\pi} 
 N(p_T,y) \sum_{n=-\infty}^{\infty} {V}_n(p_T,y)\,e^{- i n \varphi}\,, 
 \label{eq:flowdef}
\end{equation}
where we consider particles of transverse momentum $p_T$, rapidity $y$ and energy $E$, corresponding to a 
particle density $N(p_T,y)$ in momentum space.  
The azimuthal angle in momentum space is denoted by $\varphi$. 
Here, the harmonics $V_n$ are defined as complex numbers of modulus and phase corresponding 
to the magnitude and orientation of the anisotropies, respectively. 
Here, $V_n$ is normalized by the particle density in momentum space $N(p_T,y)$. 

The response of elliptic and triangular flow, $V_2$ and $V_3$, to the initial geometry 
is given, to a good approximation, by 
\begin{equation}
 V_n \simeq \kappa_n\,\epsilon_n\,,
 \label{eq:mapping0}
\end{equation} 
where the properties of the QGP are encoded in the single constant $\kappa_n$, 
and $\epsilon_n$ is an eccentricity characterizing the initial geometry, 
the precise definition of which may vary \cite{Bhalerao:2005mm,Bhalerao:2006tp,Teaney:2010vd,Gardim:2011xv,Teaney:2012ke,Gardim:2014tya,Fu:2015wba,Rao:2019vgy,Qin:2010pf,Qiu:2011iv,Niemi:2012aj,Giacalone:2017uqx,Wei:2018xpm,Sievert:2019zjr,Zhao:2020pty}. 
For $n=2$, for instance, we take 
\begin{equation}
 \epsilon_2 \equiv  - 
 \dfrac{ \{r^2\,e^{2 i\,\phi}\} - \{r\,e^{i\,\phi} \}^2  }{\{r^2\} - \{r\,e^{i\,\phi} \}\{r\,e^{-i\,\phi} \}}\,,
 \label{eq:eccdef}
\end{equation}
where $\phi$ is the azimuthal angle in position space, $r=|\vec x|$ and we define the spatial average 
\begin{equation}
 \qquad  \{ (\cdots)\} \equiv \frac{\int d^2 x \,\rho(\vec x)\, (\cdots)}{\int d^2 x\, \rho(\vec x)}\,,
 \label{eq:spatialavg}
\end{equation}
in which 
$\rho(\vec x) \equiv T^{\tau\tau} (\vec x)$, where $T^{\mu\nu}$ is the energy-momentum tensor, is the initial transverse energy density 
 in the laboratory frame, 
 at the position $\vec x$ in the transverse plane.
While the relation \eqref{eq:mapping0} is usually employed for integrated flow vectors, we here extend it to 
the differential flow $V_n(p_T)$, by considering independent values of $\kappa_n(p_T)$ in each momentum bin \cite{Wei:2019wdt}.  
An extension of Eq.~\eqref{eq:mapping0} to rapidity-dependent hydrodynamic response was considered in \cite{Franco:2019ihq,Li:2020ucq}. 

\begin{figure}[t]
 \centering
 \includegraphics[width=0.5\textwidth]{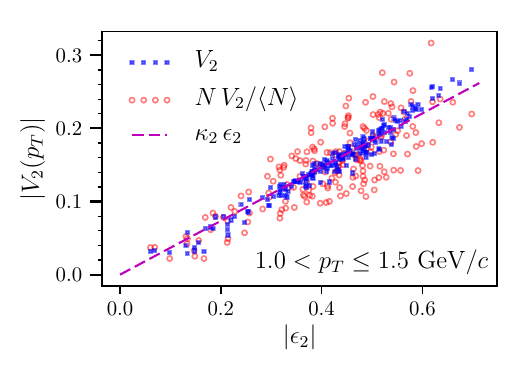}
 \caption{Scatter plot of the the elliptic flow harmonic, 
 versus the corresponding eccentricity $\epsilon_2$, in modulus, 
 for 40\%--50\% centrality $Pb+Pb$ collisions at $\sqrt{s_{NN}}=2.76$ TeV. 
 Solid blue squares correspond to values of the ``flow per particle'' $V_2(p_T)$, as defined in Eq.~\eqref{eq:flowdef}, while empty 
 red circles include the effect of multiplicity fluctuations by showing instead the combination $N(p_T)\,V_2(p_T)/\langle N(p_T) \rangle$. 
 The dashed magenta line shows the elliptic flow predicted from relation Eq.~\eqref{eq:mapping0}. 
 As expected, the flow per particle is better correlated to the geometry of the system, in comparison to the ``total flow''  $N(p_T)\,V_2(p_T)$. 
 }
 \label{fig:scatterMapping}
\end{figure}

Despite the success of Eq.~\eqref{eq:mapping0}, anisotropic flow may also respond to other
features of the initial state.  
In Fig.~\ref{fig:scatterMapping}, the solid blue squares represent values of $V_2$ and $\epsilon_2$ for a set of 
events simulated in a state-of-the-art hydrodynamic model \cite{NunesdaSilva:2020bfs,NunesdaSilva:2018viu} using  \trentonosp +\textsc{Music}+UrQMD \cite{Moreland:2014oya,Schenke:2010nt, Schenke:2011bn,Paquet:2015lta,Shen:2014vra,Bass:1998ca,Bleicher:1999xi,Bernhard:2018hnz}. 
The remarkable correlation between the two quantities visibly supports the approximation in  Eq.~\eqref{eq:mapping0} (dashed magenta line), 
indicating that elliptic flow fluctuations are mostly driven by a linear response to $\epsilon_2$. 
At the same time, the spread around linear correlation points to small corrections to the approximate linear response, 
which may originate from the finer details of the initial transverse energy distribution~\cite{Gardim:2017ruc,Kozlov:2014fqa,Noronha-Hostler:2015coa}, or from nonlinear response \cite{Gardim:2017ruc,Teaney:2012ke,Qian:2013nba,Yan:2015jma,CMS-PAS-HIN-16-018,Acharya:2017zfg,Acharya:2020taj,Wen:2020amn}. 
Extensions of  Eq.~\eqref{eq:mapping0} to contemplate such corrections were proposed and studied in Refs.~\cite{Teaney:2010vd,Gardim:2011xv,Teaney:2012ke,Gardim:2014tya,Noronha-Hostler:2015dbi,Fu:2015wba,Rao:2019vgy,Wei:2018xpm,Wei:2019wdt}.

In this paper, we set out to investigate how corrections to Eq.~\eqref{eq:mapping0}, indicated by the spread of the blue squares in Fig.~\ref{fig:scatterMapping}, 
might be experimentally studied through a principal component analysis (PCA) of anisotropic flow fluctuations \cite{Bhalerao:2014mua,Mazeliauskas:2015vea,Mazeliauskas:2015efa,Cirkovic:2016kxt,Bozek:2017thv,Gardim:2019iah,Sirunyan:2017gyb,Liu:2019jxg,Hippert:2019swu,Liu:2020ely}. 
In particular, we explore which features of the initial geometry are most relevant for understanding this analysis. 
The connection between subleading anisotropic flow and initial-state anisotropies 
was previously investigated in Refs.~\cite{Mazeliauskas:2015vea,Mazeliauskas:2015efa}, where different methods 
 were employed to interpret the original observables of Ref.~\cite{Bhalerao:2014mua}.

In Sec.~\ref{sec:PCA}, we present the PCA of  flow fluctuations to be employed in our analysis \cite{Hippert:2019swu}. 
In Sec.~\ref{sec:mapping}, we discuss a mapping of hydrodynamic response which extends  Eq.~\eqref{eq:mapping0} to encompass nonlinear response and finer details of the initial geometry \cite{Gardim:2011xv}. 
Then, we apply this mapping to simulated hydrodynamic events and employ it to understand the PCA of anisotropic flow. 
Results are presented and discussed in Sec.~\ref{sec:results}  and our main conclusions are summarized in Sec.~\ref{sec:conclusions}.

\begin{figure}[t]
 \centering
 \includegraphics[width=0.5\textwidth]{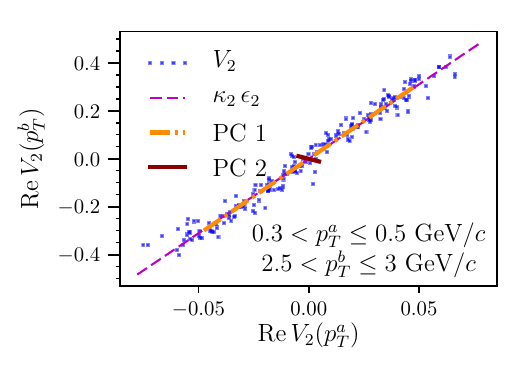}
 \caption{Scatter plot of the real part of the elliptic flow harmonic, $V_2(p_T)$, as defined in Eq.~\eqref{eq:flowdef}, in two separate 
 transverse-momentum bins $p_T^a$ and $p_T^b$ (blue squares), for 40\%--50\% centrality $Pb+Pb$ collisions at $\sqrt{s_{NN}}=2.76$ TeV. It is noticeable that the flow in the two bins 
 is strongly correlated by fluctuations of the 
 eccentricity $\epsilon_2$, as made clear by the prediction from Eq.~\eqref{eq:mapping0} (dashed magenta line). 
 Also shown are the first (``PC 1'', dot-dashed, orange line) and second (``PC 2'', solid, dark red line) principal components 
 of elliptic flow fluctuations, projected onto the subspace described by $(V_2(p_T^a),V_2(p_T^b))$. The first component 
 is associated with fluctuations of $\epsilon_2$, while the second one is related to corrections to Eq.~\eqref{eq:mapping0}.}
 \label{fig:scatterPCA}
\end{figure}

\section{Principal component analysis of flow fluctuations}
\label{sec:PCA}

We wish to find measurable consequences of corrections to relation \eqref{eq:mapping0}. 
However, only the left-hand side of this relation is accessible in experiments. 
Thus, fluctuations of the initial geometry must be inferred from fluctuations of anisotropic flow. 
One way this can be achieved is by exploring correlations between flow harmonics at different 
momentum bins. For concreteness, the blue squares in Fig.~\ref{fig:scatterPCA} display a scatter plot of the elliptic flow coefficients 
$V_2(p_T^a)$ and $V_2(p_T^b)$, measured from particles of two different bins $a$ and $b$, 
with $p_T^a\neq p_T^b$, for a set of simulated events in a hybrid event-by-event hydrodynamic model  \cite{NunesdaSilva:2020bfs,NunesdaSilva:2018viu}. 
In this figure, correlations predicted by  Eq.~\eqref{eq:mapping0} are represented by the magenta dashed line, 
of slope $\kappa_2(p_T^b)/\kappa_2(p_T^a)$. 
Once again, 
the spread of points around the linear expectation implies that fluctuations 
of anisotropic flow are not entirely determined by fluctuations of $\epsilon_n$ alone, and in fact, elliptic flow coefficients at different transverse momentum fluctuate slightly differently from one another.

 This deviation from perfect correlation in Fig.~\ref{fig:scatterPCA} can be quantified by a principal component 
analysis. In fact, this analysis can be carried out considering 
correlations among all the different transverse-momentum bins \cite{doi:10.1002/0470013192.bsa501,Bhalerao:2014mua}. 
Principal component analysis is a standard multivariate method that 
allows one to isolate linear combinations of variables which are linearly uncorrelated.  
By ordering the eigenvectors of the covariance matrix according to the eigenvalues, one can sort out 
which are the main directions of fluctuation --- or principal components --- within a given space of correlated variables \cite{doi:10.1002/wics.101}. 
Figure~\ref{fig:scatterPCA} shows the projections of the first (``PC 1'') and second (``PC 2'') principal 
components of elliptic flow onto the subspace spanned by $(V_2(p_T^a),V_2(p_T^b))$. 
The  first, or leading, component of elliptic flow lies along the expectations from Eq.~\eqref{eq:mapping0}, indicating that this component is related to fluctuations of $\epsilon_2$. On the other hand, this is not the case for the subleading component, which should be linked to other sources of fluctuation.

The PCA of anisotropic flow was first proposed in Ref.~\cite{Bhalerao:2014mua}. This original proposal was further explored 
in several papers \cite{Mazeliauskas:2015vea,Mazeliauskas:2015efa,Cirkovic:2016kxt,Bozek:2017thv,Gardim:2019iah,Liu:2019jxg,Hippert:2019swu,Liu:2020ely}  and experimentally measured by the CMS Collaboration \cite{Sirunyan:2017gyb}. 
While an event-by-event determination of the azimuthal distribution of particles, and thus of the flow harmonics $V_n(p_T)$, 
is severely hindered by the limited number of particles, a covariance matrix reflecting 
correlations among different bins can be safely extracted from two-particle correlations. 
The principal components $V_n^{(\alpha)}(p_T)$ can be found from the spectral decomposition of this matrix:
\begin{align}
\begin{split}
  \langle V_n(p_T^a)\,V_n^*(p_T^b) \rangle &= \sum_\alpha^M\lambda^{(\alpha)}\, \psi_n^{(\alpha)}(p_T^a)\, \psi_n^{(\alpha)}(p_T^b)\\
  &=  \sum_\alpha^M V_n^{(\alpha)}(p_T^a)\, V_n^{(\alpha)}(p_T^b)\,,
  \label{eq:specPCA}
\end{split}
\end{align}
where $M$ is the number of transverse-momentum bins, and $\lambda^{(\alpha)}$ and $\psi_n^{(\alpha)}(p_T)$ are the eigenvalues and eigenvectors, ordered in descending order $\lambda^{(\alpha)}\geq\lambda^{(\alpha+1)}$, 
and
\begin{equation}
  V_n^{(\alpha)}(p_T) \equiv \sqrt{\lambda^{(\alpha)}}\, \psi_n^{(\alpha)}(p_T)\,. 
\end{equation}
Because the covariance matrix is Hermitian, positive semi-definite and, assuming symmetry under parity transformations, also real, 
 $V_n^{(\alpha)}(p_T)$ can be defined as real functions of the transverse momentum.  
While a precise measurement of the covariance matrix in Eq.~\eqref{eq:specPCA} might be a challenge, 
this matrix was shown to be nearly equivalent to the alternative one introduced in \cite{Hippert:2019swu}, 
which in turn should be straightforward to measure. 

In Fig.~\ref{fig:scatterPCA}, one observes a clear hierarchy between the first or leading principal component ($\alpha=1$) --- 
corresponding to the dominant source of fluctuations --- and the much smaller subleading principal component ($\alpha=2$) --- related to subdominant 
fluctuations \cite{Bhalerao:2014mua}. 
By truncating Eq.~\eqref{eq:specPCA} at $\alpha_{\textrm{max}} \leq M$, such that $\lambda^{(\alpha_{\textrm{max}})}\ll \lambda^{(1)}$, 
one can characterize the covariance matrix, a two-variable function, by only a few functions $V_n^{(\alpha\leq\alpha_{\textrm{max}})}(p_T)$ 
of a single variable, representing the projection of the principal components upon each momentum bin.    In general, even across the entire measured momentum range, there is a strong hierarchy such that the matrix can be accurately represented by two or three principal components.
Thus, the PCA allows for an optimal, compact visualization of two-particle correlations from fluctuations of $V_n(p_T)$ \cite{Bhalerao:2014mua}.

In Eq.~\eqref{eq:specPCA}, the obtained components depend on how the spectral condition is defined. 
Here, we write the eigenvalue equation as
\begin{equation}
 \sum_{b=1}^M V_{n\Delta}(p_T^a,p_T^b)\, V_n^{(\alpha)}(p_T^b)\,W(p_T^b)\, \Delta p_T^b 
 = \lambda^{(\alpha)}\, V_n^{(\alpha)}(p_T^a)\,,
\end{equation}
where $V_{n\Delta}(p_T^a,p_T^b)\equiv \langle V_n(p_T^a)\,V_n^*(p_T^b) \rangle$ and the index $b$ 
is summed over all transverse-momentum bins, each with a weight $W(p_T^b)$ \cite{Hippert:2019swu}.  
The weight function can be chosen so as to emphasize different parts of the spectrum. 
Natural choices of weight include $W(p_T)=1$, for uniform emphasis across $p_T$, and 
$W(p_T)=\langle N(p_T) \rangle$, focusing on more occupied momentum bins. 
In this work, we adopt the former choice and take $W=1$. 

In the original proposal of Ref.~\cite{Bhalerao:2014mua}, the covariance matrix of the ``total flow''  was considered. That is, 
$\mathcal{V}_n\equiv N(p_T,y)\,V_n(p_T,y)$ at each momentum bin, and the flow vectors  were not normalized by multiplicity.  
However, a covariance matrix of the ``flow per particle'' $V_n(p_T,y)$ is better suited to our needs \cite{Bozek:2017thv, Hippert:2019swu}. 
In fact, in \cite{Hippert:2019swu}, an important difference between subleading fluctuations of the ``total'' and the 
``per-particle'' anisotropic flow was found. 
This difference is clearly visible in Fig.~\ref{fig:scatterMapping}, where empty red circles represent values of $N\,V_2/\langle N\rangle$ 
in different events. Fluctuations of $V_2$ correlate better with the geometry of the events, while fluctuations of $N\,V_2$ are 
affected by fluctuations of particle number \cite{Hippert:2019swu}. 
It is noteworthy that the deviation is larger for higher values of $|\epsilon_2|$, corresponding to more peripheral 
collisions, where multiplicity fluctuations are more important.

The decorrelation among flow fluctuations at different values of the momentum can also be explored using the factorization breaking coefficient 
$r_n(p_T^a,p_T^b)$ \cite{Gardim:2012im,Heinz:2013bua,Kozlov:2014fqa,Gardim:2017ruc,Kozlov:2014hya,Shen:2015qta,Zhao:2017yhj,Bozek:2018nne,Khachatryan:2015oea,CMS:2013bza,Acharya:2017ino}. 
However, this approach reveals the importance of subleading fluctuations only \emph{in relative terms}. 
In case the dominant flow fluctuations stemming from eccentricity fluctuations become too large, as is the case for peripheral collisions, 
subleading flow fluctuations will only weakly impact the value of $r_n(p_T^a,p_T^b)$.

\section{Mapping hydrodynamic response} 
\label{sec:mapping}

Having built some intuition on the PCA of anisotropic flow, we now turn to a more quantitative study of 
its precise physical content. 
More specifically, we aim at determining which features of the fluctuating initial geometry are essential to 
the second principal component. 
To that end, we employ an approach based on Refs.~\cite{Gardim:2011xv,Gardim:2014tya}, explained below.  

Let us assume that the QGP evolves deterministically, starting from early times, $\tau \lesssim \tau_0$. 
The energy-momentum tensor $T^{\mu\nu}(\tau\geq \tau_0,\vec x)$ at later times 
is, thus, fully determined by its components at $\tau = \tau_0$. 
As a consequence, the final single-particle distribution is a functional of $T^{\mu\nu}(\tau_0,\vec x)$:\footnote{We here assume boost-invariant initial conditions. 
A study of hydrodynamic response beyond 2+1-dimensional hydrodynamics can be found in \cite{Franco:2019ihq,Li:2020ucq}.} 
\begin{align}
\dfrac{d N}{dy \,p_T dp_T\, d\varphi} &= \mathcal{F}[T^{\mu\nu}(\tau_0,\vec x)]\,.
\label{eq:PartDistTmunu}
\end{align}

Our purpose is to model the azimuthal dependence of $\mathcal{F}[T^{\mu\nu}(\tau_0,\vec x)]$ in a systematic manner. 
This can be achieved by employing a cumulant expansion of the initial conditions to define 
\emph{eccentricities} $\epsilon_{n,m}$. 
Thus, one can establish phenomenological relations $V_n \approx \mathcal{F}_n[\{\epsilon_{n',m'}\}]$, where 
$\mathcal{F}_n$ can be approximated by a power series in $\epsilon_{n',m'}$. 
This series is restricted to terms with the correct symmetries and ordered according to a hierarchy of scales, in a Ginzburg-Landau fashion. 
As will become clear, the leading lowest-order term in such a series,  $\propto\epsilon_{n,n}$, is related to the usual eccentricity scaling 
of Eq.~\eqref{eq:mapping0}, while corrections give rise to the subleading principal components of anisotropic flow.

\subsection{Characterizing the initial geometry}

For simplicity, we assume a ``static'' transverse energy distribution at $\tau = \tau_0$ and neglect components of 
$T^{\mu\nu}(\vec x, \tau_0)$ other than the energy density $\rho(\vec x) \equiv T^{\tau\tau}(\tau_0,\vec x)$.\footnote{A similar treatment 
including other components of $T^{\mu\nu}$ can be found in Ref.~\cite{Sousa:2020cwo}. } 
It proves useful to take its Fourier transform,  
\begin{equation}
 \rho(\vec k) = \int d^2 x \, \rho(\vec x)\, e^{i \vec k \cdot \vec x}\,,   
 \label{eq:rhofourier}
\end{equation}
so that different values of $|\vec k|$ probe $\rho(\vec x)$ at different scales. 
In fact, $\rho(\vec k)$ can be interpreted as a moment generating function, from which eccentricities might be extracted \cite{Teaney:2010vd}. 
Because of their transformation properties under rotation, 
it is convenient to define $z\equiv x + i\, y$ and $k_z \equiv k_x + i\,k_y$. 
Moments of $z$ and $z^*$ are given by 
\begin{equation}
\{z^j\,{z^*}^\ell\} = \dfrac{\left(-2\,i\right)^{j+\ell}}{\rho_0}\,\frac{\partial^{j+\ell}\rho(\vec k)}{\partial {k_z^*}^j \partial k_z^\ell}\bigg|_{k =0}\,,
 \label{eq:momdef}
\end{equation}
where $\rho_0 \equiv \rho(\vec k =\vec 0)$ and we use the definition in Eq.~\eqref{eq:spatialavg}.

The moments 
$\{z^i\,{z^*}^j\}$ are not invariant under translations and depend on the choice of 
coordinate system \cite{Gardim:2014tya}. This is related to the fact that $|k_x|^{-1}$ and $|k_y|^{-1}$ are actually scales of 
distance to an arbitrary origin, not of separation between points.  
This issue can be solved by using, instead, the function
\begin{equation}
 W(\vec k) \equiv \log\left(\rho(\vec k)/\bar\rho\right)\,, 
 \label{eq:Wdef}
\end{equation}
where $\bar \rho$ sets an arbitrary scale with the same units as $\rho(\vec k)$. 
Notice that under a translation, 
\begin{equation}
 \vec x \to \vec x + \vec d:\;\; 
 W(\vec k) \to W(\vec k) + i\,\vec k\cdot \vec d\,, 
 \label{eq:Wtranslation}
\end{equation}
so that all but the first derivatives of $W(\vec k)$ are invariant under translations. 
Cumulants can be computed from  
\begin{equation}
\{z^j\,{z^*}^\ell\}_{\textrm{cml}} \equiv \left(-2\,i\right)^{j+\ell}\,\frac{\partial^{j+\ell}W(\vec k)}{\partial {k_z^*}^j \partial k_z^\ell}\bigg|_{k =0}\,. 
\label{eq:cumdef}
\end{equation}

To study transformation properties under rotations, it is  convenient to employ polar coordinates, where 
$z \equiv r\,e^{i \phi}$. 
We thus define 
\begin{align}
 \rho_{j-\ell,j+\ell} &\equiv  \{z^{j}\,{z^*}^{\ell}\} =\{r^{j+\ell}\,e^{i (j-\ell)\phi}\}\,,
 \label{eq:rhonm}\\
 W_{j-\ell,j+\ell} &\equiv \{z^{j}\,{z^*}^{\ell}\}_{\textrm{cml}} = \{r^{j+\ell}\,e^{i (j-\ell)\phi}\}_{\textrm{cml}} \,,
 \label{eq:Wnm}
\end{align}
which transform as the harmonics $V_n$ 
under rotations, 
\begin{equation}
  \phi \to \phi + \delta:\;\; \rho_{n,m} \to \rho_{n,m}\,e^{i n \, \delta}\,,\;\; W_{n,m} \to W_{n,m}\,e^{i n \, \delta}\,.
\label{eq:Wrotation}
  \end{equation}
Because of Eqs.~\eqref{eq:rhonm} and \eqref{eq:Wnm},
 $\rho_{n,m}$ and $W_{n,m}$ are defined only for even, non-negative values of $m-|n|$. They are taken to vanish otherwise. 
 
The set of all $\rho_{n,m}$, or all $W_{n,m}$, is sufficient to fully recover the shape of the initial condition $\rho(\vec x)$. 
In fact, expanding $\rho(\vec k)$ and $W(\vec k)$ in powers of $k_z$ and $k_z^*$, and using Eqs.~\eqref{eq:momdef} and \eqref{eq:cumdef}, one finds 
that these moments and cumulants can be interpreted as series coefficients 
\cite{Teaney:2010vd}:
\begin{align}
\rho(\vec k ) &= \rho_0\,\sum_{m=0}^\infty \sum_{n=-m}^{m} \dfrac{(i/2)^m\, \rho_{n,m}}{\left(\frac{m+n}{2}\right)!\,\left(\frac{m-n}{2}\right)!}k^m\,e^{-i n \phi_k}\,,
\label{eq:seriesrho}
\\
W(\vec k ) &= \sum_{m=0}^\infty \sum_{n=-m}^{m} \dfrac{(i/2)^m\, W_{n,m}}{\left(\frac{m+n}{2}\right)!\,\left(\frac{m-n}{2}\right)!}k^m\,e^{-i n \phi_k}\,,
\label{eq:seriesW}
 \end{align}
where it becomes clear that larger values of $m$ become important at higher values of $k$ and, thus, at smaller spatial scales. 
Also, if cumulants with $m\geq 3$ are neglected, one obtains a simple Gaussian distribution. 
Notice as well that only coefficients with positive $n$ are required, because 
 $\rho_{-n,m} = \rho_{n,m}^*$ and $W_{-n,m} = W_{n,m}^*$. 

More details, including explicit, general expressions for the cumulants $W_{n,m}$ in terms of the moments $\rho_{n,m}$ can be found in Appendix \ref{app:cumsandmoms}. 
As an example, we write down the expressions for $m=2$:
\begin{align}
W_{0,2} &=
\{r^2\} -  \{r\, e^{-i\,\phi_x}\}\, \{r\, e^{i\,\phi_x}\}
\label{eq:W02}
\,,\\
W_{2,2} &=
\{r^2\, e^{2i\,\phi_x}\} -  \{r\, e^{i\,\phi_x}\}^2 
\label{eq:W22}
\,,
\end{align}
and $m=3$:
\begin{align}
 \begin{split}
 W_{1,3}  ={}& 
  \left\{ r^3\,e^{i\phi_x} \right\}   -   \left\{ r^2\,e^{2i\phi_x} \right\} \, \left\{ r\,e^{-i\phi_x} \right\}\\
   &-    2\,\left\{ r^2 \right\} \, \left\{ r\,e^{i\phi_x} \right\}      +    2\,\left\{ r\,e^{i\phi_x} \right\}^2 \, \left\{ r\,e^{-i\phi_x} \right\}  
\,,
\label{eq:W13}
\end{split}
\\
\begin{split}
 W_{3,3}  ={}& \left\{ r^3\,e^{3i\phi_x} \right\}   -  3\, \left\{ r^2\,e^{2i\phi_x} \right\} \, \left\{ r\,e^{i\phi_x} \right\}\\
    &+    2\,\left\{ r\,e^{i\phi_x} \right\}^3
\,.
\label{eq:W33}
\end{split}
\end{align}

\subsection{Hydrodynamic response to initial geometry}
\label{sub:mappresponse}

The cumulant expansion above allows one to characterize the initial 
geometry of the system with a set of complex numbers $W_{n,m}$. 
The index $n$ specifies a harmonic of the azimuthal distribution of energy, 
while $m$ indirectly determines the length scales contributing to each cumulant, as well as 
the scaling with the typical transverse size $L$:
\begin{equation}
 W_{n,m} = \{r^m\,e^{in\phi}\}_{\textrm{cml}} \propto L^m\,.
 \label{eq:cmlsizescaling}
\end{equation}

Because of the oscillating exponential $e^{in\phi}$, realistic initial conditions are expected to have 
$W_{n\neq 0,m} \ll L^m$. Thus, we can characterize the initial-state anisotropies with typically 
small, dimensionless, system-size independent eccentricities \cite{Gardim:2011xv,Gardim:2014tya}:
\begin{equation}
 \epsilon_{n,m} \equiv -\dfrac{W_{n,m}}
  {R^m}\,,
 \label{eq:defecc}
\end{equation}
where we use $R = \sqrt{W_{0,2}}$ as a measure of system size.
In addition to having well defined rotational symmetries, all eccentricities except $\epsilon_{\pm1,1}$ are 
invariant under translation, as can be seen from Eq.~\eqref{eq:Wtranslation}.

From Eq.~\eqref{eq:PartDistTmunu}, $V_n(p_T)$ is a function of the eccentricities in Eq.~\eqref{eq:defecc}, 
which we can expand as a power series. This power series is restricted to terms which transform 
as $V_n$ under rotations, which considerably simplifies its form. 
Up to linear response, we have:
\begin{equation}
 V_n(p_T) \approx \sum_{\substack{m=n\\m\neq 1}}^{m_{\textrm{max}}}\,\kappa^{(n)}_m(p_T)\, \epsilon_{n,m} + \mathcal{O}(\epsilon_{n,m_{\textrm{max}}})+ \mathcal{O}(\epsilon^2)\,,
 \label{eq:VnEccsLin}
 \end{equation}
where, assuming that larger scales contribute the most, we neglect eccentricities with $m>m_{\textrm{max}}$. 
By enforcing translational invariance, we have excluded $m=1$ from the series expansion. 
Including higher powers of $\epsilon_{n,m}$, up to $p_{\textrm{max}}$, we find 
\begin{multline}
 V_n(p_T) \approx 
  \sum_{p=1}^{p_{\textrm{max}}}\, \sum_{\{n',m'\}}^{\sum n'_i=n}\, \kappa^{(n)}_{\{n',m'\}}(p_T)\,\prod_{i=1}^{p} \epsilon_{n'_i,m'_i} +\\
 +\mathcal{O}(\epsilon_{n,m_{\textrm{max}}})+ \mathcal{O}(\epsilon^{p_{\textrm{max}}+1})\,,
 \label{eq:VnEccsNonlin}
\end{multline}
where the sum includes negative values of $n'$ and is restricted to terms with the correct rotational and translational symmetries. 

\begin{figure*}
 \centering
 \includegraphics[width=\textwidth]{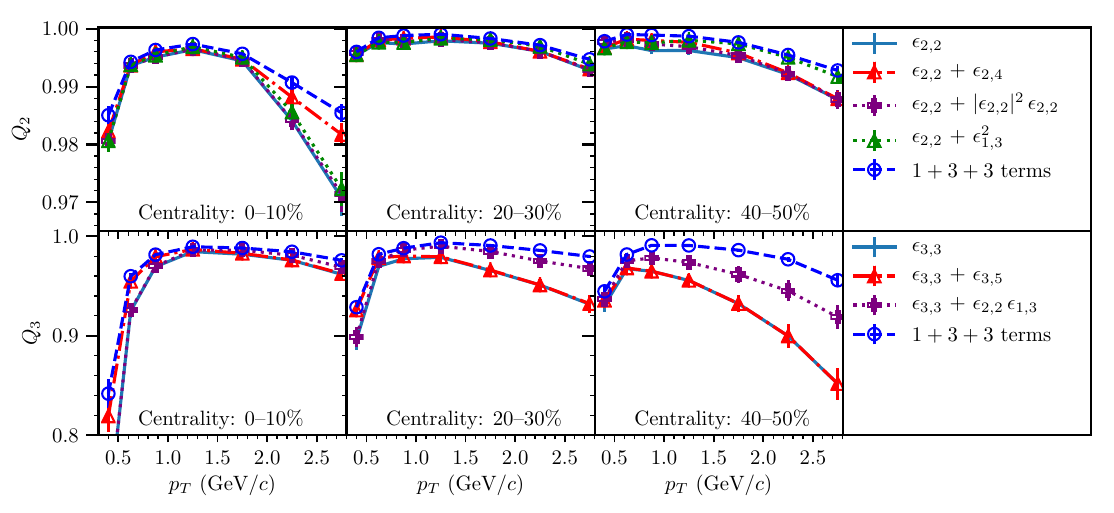}
 \caption{ Pearson correlation coefficient between the flow harmonics $V_2(p_T)$ (upper panel) and $V_3(p_T)$ (lower panel) and predictions of their event-by-event fluctuations from eccentricities of the 
 initial transverse geometry. Different curves correspond to different predictors, with the blue dashed curve corresponding to the full expressions in Eqs.~\eqref{eq:predV2} and \eqref{eq:predV3}. 
 Events are simulated for $Pb+Pb$ collisions at $\sqrt{s_{NN}}=2.76$ TeV, within a hybrid event-by-event hydrodynamic model (\trentonosp +\textsc{Music}+UrQMD). }
 \label{fig:Pearson23}
\end{figure*}

The coefficients $\kappa(p_T)$ in Eqs.~\eqref{eq:VnEccsLin} and \eqref{eq:VnEccsNonlin} are responsible for encoding 
all the information on the relevant QGP properties, e.g. 
equation of state and transport coefficients. 
They can be obtained by minimizing the squared norm of the residuals \cite{Gardim:2011xv,Gardim:2014tya}
\begin{equation}
 \delta_n = V_n^{(\textrm{hydro})}[\rho_0(\vec x)] -  V_n^{(\textrm{est})}(\{\epsilon,\kappa\})
\end{equation}
in each transverse-momentum bin, where $V_n^{(\textrm{hydro})}$ and $V_n^{(\textrm{est})}$ are the flow harmonics from 
full hydrodynamic simulations and estimates obtained from the power series in Eq.~\eqref{eq:VnEccsNonlin}, respectively. 
By taking the derivative of $\langle|\delta_n|^2\rangle$ with respect to $\kappa$, we arrive at
 the system of equations
\begin{equation}
\sum_{\{n',m'\}}^{\sum n'_i=n} \operatorname{Re}\,\langle \varepsilon_{\{n,m\}}^*\varepsilon_{\{n',m'\}}\rangle\kappa^{(n)}_{\{n',m'\}} 
= \operatorname{Re}\, \langle V_n^*\, \varepsilon_{\{n,m\}}\rangle\,,
\label{eq:optkappa}
\end{equation}
where 
$\varepsilon_{\{n,m\}} \equiv\prod \epsilon_{n_i,m_i}$. 
Solving Eq.~\eqref{eq:optkappa} yields optimal values of $\kappa$, which can be employed to predict the flow harmonic $V_n(p_T)$.
Any dependence of the final flow harmonics 
on $W_{0,m}$ must be incorporated in the coefficients $\kappa$, which for this reason 
are mildly centrality dependent. 

We emphasize that this prescription treats each momentum bin independently.  As such, a description of correlated fluctuations between different momentum bins, as measured with PCA, is a non-trivial test of the framework.  

Both Eqs.~\eqref{eq:VnEccsLin} and \eqref{eq:VnEccsNonlin} are generalizations of Eq.~\eqref{eq:mapping0}. 
Similar expressions have been presented in Refs.~\cite{Teaney:2010vd,Teaney:2012ke,Gardim:2011xv,Gardim:2014tya}.
Unlike previous approaches, however, we here undertake the description of the \emph{differential} flow harmonics 
$V_n(p_T)$, by also promoting the coefficients $\kappa$ to functions of $p_T$. 
We  stress that the eccentricities $\epsilon_{n,m}$ do not depend on the transverse momentum of the particles in any way, being determined solely by the initial transverse energy-density profile $T^{\tau\tau}(\vec x)$ at $\tau=\tau_0$. Thus, the transverse-momentum dependence of the flow harmonics $V_n(p_T)$ is fully encoded in the response coefficients. Response coefficients for the leading and subleading terms in two-term predictors of triangular and elliptic flow can be found in Appendix~\ref{app:coeffs}. 

Because the flow harmonics can be measured in experiments, while the eccentricities are available in models for the initial conditions, the relation in Eq.~\eqref{eq:mapping0} was used to extract information on the response coefficients of the QGP, $\kappa_n$. This can be achieved by comparing measurements of the flow harmonics $V_n$ to model calculations of the initial eccentricities $\epsilon_n$ \cite{ALICE:2011ab}. However, in Eqs.~\eqref{eq:VnEccsLin} and \eqref{eq:VnEccsNonlin}, the presence of multiple terms renders the direct extraction of information from experimental data less straightforward. Even for a small number of terms, such an extraction would most likely require extra information on the event-by-event probability distribution for the flow harmonics, as extracted, for instance, from the PCA itself or from the unfolding approach \cite{ATLAS-CONF-2012-049,Jia:2013tja}.

\begin{figure*}
 \centering
 \includegraphics[width=\textwidth]{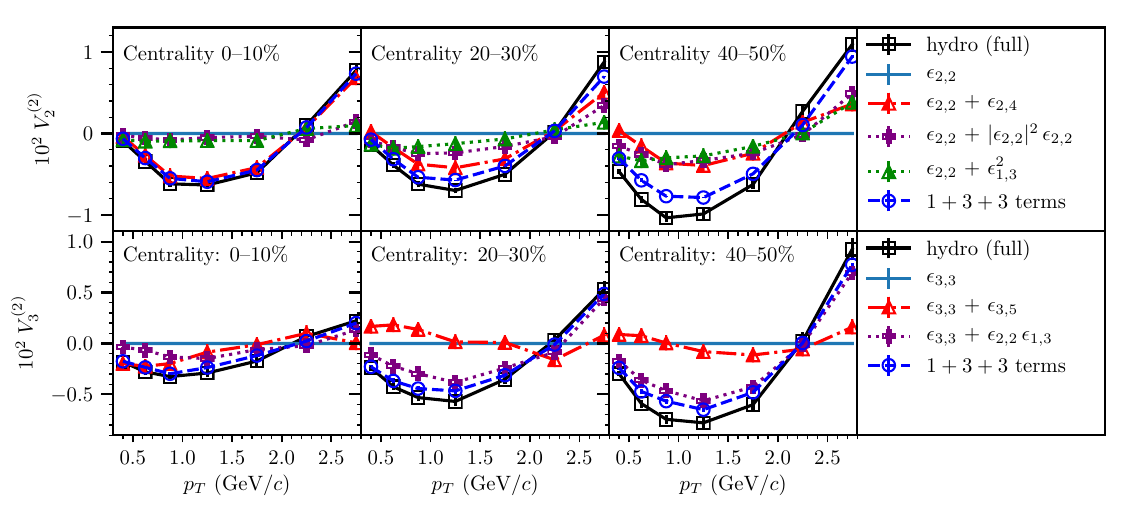}
 \caption{ Second principal component of elliptic (upper panel) and triangular (lower panel) flow, both from full hydrodynamic simulations and event-by-event predictions from eccentricities of the 
 initial geometry. Different curves correspond to different predictors, with the black solid curve corresponding to the full hydrodynamic results and the blue dashed curve corresponding to the full 
 expressions in Eqs.~\eqref{eq:predV2} and \eqref{eq:predV3}. 
 Events are simulated for $Pb+Pb$ collisions at $\sqrt{s_{NN}}=2.76$ TeV, within a hybrid event-by-event  hydrodynamic model (\trentonosp +\textsc{Music}+UrQMD).  }
 \label{fig:PCAmapp23}
\end{figure*}

\section{Results} 
\label{sec:results}

We apply the mapping of Section~\ref{sec:mapping} to $Pb+Pb$ collisions at center-of-mass energy 
$\sqrt{s_{NN}} = 2.76$ TeV simulated in an event-by-event hybrid model --- the same simulated events 
shown in Figs.~\ref{fig:scatterMapping} and \ref{fig:scatterPCA} \cite{NunesdaSilva:2020bfs,NunesdaSilva:2018viu}. 
Our boost-invariant initial conditions are generated with the parametric model \trento \cite{Moreland:2014oya} and fed into 
relativistic viscous hydrodynamics as implemented in \textsc{Music}  \cite{Schenke:2010nt,Schenke:2011bn,Paquet:2015lta}. 
Model parameters 
for \trento (except for the normalization factor) and for the parametrization of the hydro viscosities  
are taken from the Bayesian analysis of \cite{Bernhard:2018hnz}, where they were optimized to describe LHC data. 
More details and results from this model can be found in Refs.~\cite{NunesdaSilva:2020bfs,Hippert:2019swu,NunesdaSilva:2018viu}. 
Hadrons are sampled from the freeze-out hypersurface using \texttt{iSpectraSampler} (iSS) \cite{Shen:2014vra} and  
their interactions in the hadron gas phase are described with UrQMD \cite{Bass:1998ca,Bleicher:1999xi}. 
A direct event-by-event determination of $V_n(p_T)$, which would be otherwise impractical, is enabled by applying an oversampling procedure,   
in which the freeze-out hypersurface of each hydrodynamic event is converted into particles multiple times, until a threshold 
number of particles is achieved. 
This artificial increase of the number of particles also has the advantage of dissolving correlations from hadronic interactions and resonance decays.

Our aim is to understand which terms and eccentricities in Eq.~\eqref{eq:VnEccsNonlin} provide the most important corrections 
to Eq.~\eqref{eq:mapping0}. In particular, we are interested in understanding the importance of linear response to higher-order eccentricities 
such as $\epsilon_{2,4}$. 
We start by writing 7-term predictors for $V_2$ and $V_3$, with three subdominant linear terms and three nonlinear terms:  
\begin{align}
\begin{split}
    V_2 \simeq {}& \kappa^{(2)}_2\,\epsilon_{2,2}  
 +  \kappa^{(2)}_4\,\epsilon_{2,4} +\kappa^{(2)}_6\,\epsilon_{2,6}  +\kappa^{(2)}_8\,\epsilon_{2,8} + \mathcal{O}(m=10)\\
 &+    \kappa^{(2)}_{(2,2)^3}\,|\epsilon_{2,2}|^2\epsilon_{2,2} +   \kappa^{(2)}_{\substack{(2,2)\\(4,4)}}\,\epsilon_{4,4}\epsilon_{2,2}^*  
  +   \kappa^{(2)}_{(1,3)^2}\,\epsilon_{1,3}^2 \\
  &+ \ldots  
  + \mathcal{O}(\epsilon^3)\,,
  \label{eq:predV2}
\end{split}
\end{align} 
\begin{align}
 \begin{split}
  V_3 \simeq {}& \kappa^{(3)}_3\,\epsilon_{3,3} 
  +  \kappa^{(3)}_5\,\epsilon_{3,5} +\kappa^{(3)}_7\,\epsilon_{3,7}  +\kappa^{(3)}_9\,\epsilon_{3,9}+ \mathcal{O}(m=11)\\
  &+  \kappa^{(3)}_{\substack{(2,2)\\(1,3)}}\,\epsilon_{2,2}\epsilon_{1,3} +  \kappa^{(3)}_{\substack{(2,2)^2\\(1,3)}}\epsilon_{2,2}^2\epsilon_{1,3}^*
  + \kappa^{(3)}_{\substack{(4,4)\\(1,3)}}\,\epsilon_{4,4}\epsilon_{1,3}^* \\  &+ \ldots
  + \mathcal{O}(\epsilon^3)\,,
 \label{eq:predV3}    
 \end{split}
\end{align}
where the only cubic terms are those involving $\epsilon_{2,2}$, expected to be the largest eccentricity \cite{Noronha-Hostler:2015dbi}. 
To understand the role of each term in Eqs.~\eqref{eq:predV2} and \eqref{eq:predV3}, 
we calculate, for different combinations of terms, predictions for the event-by-event  $V_2(p_T)$ and $V_3(p_T)$, 
which we then compare to full hydrodynamic simulations. 
The coefficients $\kappa(p_T)$ are fixed at their optimal values for each set of terms. 
A more comprehensive study, including results for $V_4(p_T)$ and $V_5(p_T)$, can be found in Appendix~\ref{app:otherharms}.

\subsection{Quality of the predictors}
\label{sec:predQ}

The quality of the estimators in Eq.~\eqref{eq:VnEccsLin} and \eqref{eq:VnEccsNonlin} 
can be assessed by measuring how they correlate with $V_n(p_T)$, at the optimum values of $\kappa$. 
To measure the degree of correlation 
we employ the Pearson correlation coefficient \cite{Gardim:2011xv,Gardim:2014tya}
\begin{equation}
 Q_n(p_T) \equiv \dfrac{ \operatorname{Re}\,\langle {V_n^*}^{(\textrm{hydro})}(p_T)\,V_n^{(\textrm{est})}(p_T) \rangle}
 {\sqrt{\langle |{V_n}^{(\textrm{hydro})}(p_T)|^2\rangle\, \langle |V_n^{(\textrm{est})}(p_T)|^2 \rangle}}\,,
\end{equation}
which is always between $-1$, corresponding to perfect anticorrelation, and $1$, corresponding to perfect correlation. 

Results for the correlation coefficients $Q_2$ and $Q_3$ are displayed in Fig.~\ref{fig:Pearson23}, for predictors constructed from different sets of terms. 
We first note that, as long as the leading eccentricity $\epsilon_{n,n}$ is included, all combinations of terms provide good predictors of $V_n$, with $Q_n$ 
consistently close to $1$. 
However, it is also visible that including new eccentricities and terms can improve the correlation even further, especially for noncentral events and higher $p_T$. 
In particular, we highlight that the full predictors in Eqs.~\eqref{eq:predV2} and \eqref{eq:predV3}  (dashed blue curve), containing seven terms, are consistently 
closer to the full simulation results, hinting at the convergence of the double expansion in Eq.~\eqref{eq:VnEccsNonlin}. 
Our predictors have their worst performance for central collisions and low transverse momentum, where 
eccentricities are expected to be smaller.

Different terms in the mapping of hydrodynamic response become more important depending on centrality and transverse momentum. 
It is to be expected that, in more peripheral collisions, increasing eccentricities render nonlinear terms more important. 
Furthermore, if larger values of $p_T$ are associated to stronger pressure gradients, it should come as no surprise that nonlinear hydrodynamic 
response becomes more important as $p_T$ increases. 
In fact, for $V_2$, the subleading linear term $\epsilon_{2,4}$ is especially important in more central collisions, while the nonlinear term $\epsilon_{1,3}^2$ becomes more 
important at higher centralities. The cubic term $|\epsilon_{2,2}|^2\epsilon_{2,2}$ --- often assumed to be the most important subdominant term in peripheral events \cite{Noronha-Hostler:2015dbi} ---
is typically less important than $\epsilon_{1,3}^2$.  It becomes relevant for non-central collisions and at low $p_T$, where most of the particles are, but also where its effect is barely visible. 
For $V_3$, the nonlinear term proportional to $\epsilon_{2,2}\epsilon_{1,3}$ also becomes more important in more peripheral events, but the $p_T$ dependence is 
even stronger. The linear term proportional to $\epsilon_{3,5}$ becomes less relevant at higher $p_T$, being eclipsed by $\epsilon_{2,2}\epsilon_{1,3}$ even in central 
collisions. The response coefficients for the leading and subleading terms in Fig.~\ref{fig:Pearson23} can be found in Appendix~\ref{app:coeffs}. 

As a caveat, we stress that the higher cumulants $W_{n,m}$ are, in general, nonlinear on the moments $\rho_{m,n}$, and the terms that we consider to be linear here might be viewed as 
nonlinear elsewhere --- if eccentricities are defined from $\rho_{m,n}$.   This is especially important for understanding the higher harmonic results presented in Appendix \ref{app:otherharms}.

\subsection{PCA from hydrodynamic response}
\label{sec:mappPCA}

In Sec.~\ref{sec:PCA}, we have argued that corrections to the usual eccentricity scaling in Eq.~\eqref{eq:mapping0}
are revealed by subleading principal components of the flow harmonics. 
By employing the results of Sec.~\ref{sec:mapping}, we can now make our argument quantitative and 
investigate how different terms and eccentricities in Eq.~\eqref{eq:VnEccsNonlin} affect the 
PCA of anisotropic flow. 
In addition to verifying our claim, we are also able to 
reveal the information contained in subleading principal components at different centralities and transverse momenta for both
 elliptic and triangular flow. 

It can be shown that the leading principal component is nearly unaffected by subleading terms, at least for $n=2,3$ \cite{Hippert:2020sat}. 
Because the third principal component of anisotropic flow can be quite small, we thus focus on the second principal component $V_n^{(2)}(p_T)$. 
Results are shown in Fig.~\ref{fig:PCAmapp23}, where we compare the subleading principal components extracted from full hydrodynamic simulations in our model (solid black curve) 
and from the predictors in Eqs.~\eqref{eq:predV2} and \eqref{eq:predV3} (dashed blue curve). 
The vertical axis shows the projection $V_n^{(2)}(p_T)$ of the second principal component on a given momentum bin, while the horizontal axis corresponds to increasing values of $p_T$. 
A good agreement is found, especially for more central collisions. 

We stress that a nonvanishing subleading principal component can only be predicted if more than one term is included in the predictor for $V_n$.  
In Fig.~\ref{fig:PCAmapp23}, this is illustrated by the flat pale-blue solid lines, corresponding to predictions from a single eccentricity $\epsilon_{n,n}$, in which case $V_n$ at each $p_T$ must fluctuate identically.  Thus, subleading PCA modes uniquely isolate higher corrections in the cumulant series.
Predictions for $V_n^{(2)}(p_T)$ including one subleading term for each harmonic are also presented. 
Once again, we find that nonlinear terms are more important for more peripheral collisions, where eccentricities become larger. 
For $V_2^{(2)}(p_T)$, the linear subleading term $\propto \epsilon_{2,4}$ provides the dominant contribution in central collisions, indicating a sensitivity to smaller scale structure of the initial state.
In more peripheral centralities, both cubic and quadratic terms are equally important. 
For $V_3^{(2)}(p_T)$, we find nonlinear terms to be more important in general. 
Already at $0-10\%$ centrality, a competition between linear and nonlinear terms is found, with the linear $\propto \epsilon_{3,5}$ term 
dominating at low $p_T$, while the nonlinear $\propto \epsilon_{1,3}\epsilon_{2,2}$ term dominates at higher $p_T$. 
For more peripheral collisions, the prediction from the linear subdominant term $\propto\epsilon_{3,5}$  
looks qualitatively different from the full simulation results --- 
it crosses the horizontal axis two times, more than required by orthogonality with $V_3^{(1)}$. 

We have also checked that the predictors in Eqs.~\eqref{eq:predV2} and \eqref{eq:predV3} provide a reasonable description of the 
third principal components of both elliptic and triangular flow. 
This confirms that the good description of subleading principal components is not fortuitous, and indicates that these predictors provide a surprisingly detailed 
description of flow fluctuations. 
Results can be found in Appendix~\ref{app:otherharms}. 

\subsection{Granularity of the initial state}
\label{sec:granular}

\begin{figure*}[t]
 \centering
 \includegraphics[width=\textwidth]{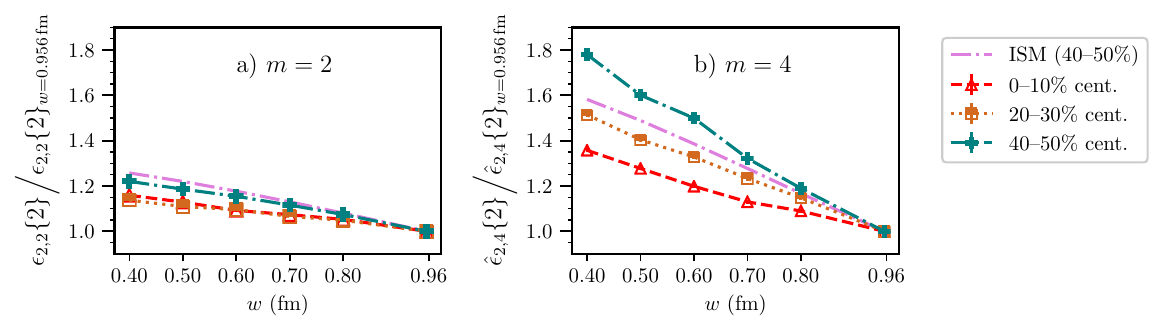}
 \caption{Scaling of fluctuations of a) $\epsilon_{2,2}$, with $m=2$, and b)  the linearly uncorrelated eccentricity $\hat\epsilon_{2,4}$, with $m=4$, as functions the nucleon-width parameter $w$. 
 Results were generated by running \trento initial conditions \cite{Moreland:2014oya} with different values of $w$, 
 a proxy for  the granularity of the initial state. 
 Thick curves correspond to results from different centrality classes.  
 Faint magenta curves correspond to the estimate in Eq.~\eqref{eq:eccgranular-Rd}, from an independent source model (ISM) with isotropic sources, for $40-50\%$ centrality. 
 }
 \label{fig:Eps24Sigma}
\end{figure*}

From Fig.~\ref{fig:PCAmapp23}, we find the subleading component of elliptic flow $V_2^{(2)}(p_T)$ to be especially sensitive to $\epsilon_{2,4}$, being dominated by 
its contributions in central collisions.
Unlike the more familiar $\epsilon_{2,2}$, this eccentricity characterizes the initial state at smaller length scales, making $V_2^{(2)}(p_T)$ a promising probe 
of the granularity of the initial transverse energy distribution \cite{Gardim:2017ruc,Kozlov:2014fqa,Noronha-Hostler:2015coa}. 

An important question is then, exactly how sensitive is the subleading PCA mode to the small-scale structure of the initial state, and what specific length scales can be probed?

To address these questions,
we first employ a simple independent source model (ISM).\footnote{ See, for instance, 
Ref.~\cite{Bhalerao:2011bp}.} 
In this model, $N$ identical sources are randomly (and independently) distributed in the transverse plane according to some probability distribution $p(\vec x)$.   
Each event then has a density
\begin{align}
\rho(\vec x) &= \sum_{i=1}^N \rho_{S}(\vec x - \vec x_i),
\end{align}
where $\rho_{S}(\vec x - \vec x_i)$ is the density distribution of a single source.\footnote{The source distribution $\rho_{S}$ can also fluctuate, with results unchanged, as long as the distribution is statistically independent of the position.}
Because $W_{n,m}$ are cumulants, the effect of the random source positions separates from the effect of the shape of each source $\rho_{S}$.  That is, we can write the single event density as a convolution 
\begin{align}
\rho(\vec x) &=\int d^2 x'\,\rho_D(\vec x')\, \rho_S(\vec x - \vec x')
\label{eq:ISMconv}
\end{align}
with
\begin{equation}
\rho_D(\vec x) \equiv \sum_{i=1}^N \delta^{(2)}(\vec x - \vec x_i)\,.
\end{equation}

Because Eq.~\eqref{eq:ISMconv} is a convolution and the generating function $W(\vec k)$ is the logarithm of a Fourier transform, contributions 
from $\rho_D$ and $\rho_S$ combine additively 
and so do all cumulants: 
\begin{align}
W(\vec k) &=  \ln \left(\frac{\rho_D(\vec k)}{\rho_D(\vec 0)}\right)
+ \ln \left(\frac{\rho_S(\vec k)}{\rho_S(\vec 0)}\right)\\
&\equiv W^D(\vec k) + W^S(\vec k)\,, \\
\implies  W_{n,m} &= W_{n,m}^D + W_{n,m}^S\,,
\end{align}
where, from now on, the superscripts $D$ and $S$ denote the distribution of sources and the average source shape, respectively.  
Assuming isotropic sources, $W_{n,m} = W_{n,m}^D$ for all $n \neq 0$, but 
the system size is still affected by the r.m.s. source radius $R_S\equiv \sqrt{W_{0,2}^S}$ \cite{Bhalerao:2011bp}, which sets the scale for  the granularity of the system :
\begin{equation}
 W_{0,2} =  R_D^2 + R_S^2 \equiv R^2\,.
\end{equation}
Thus, from Eq.~\eqref{eq:eccdef}, and assuming $R_S\ll R_D$:
\begin{align}
  \epsilon_{n,m} 
  &= \frac{R_D^m \ \epsilon_{n,m}^D}{R^m}\\
  &= \dfrac{(R^2 - R_S^2)^{m/2}\,\epsilon_{n,m}^D }{R^m}
  \label{eq:eccgranular-ex}
\\
 &\approx \left(1 - \dfrac{m}{2} \dfrac{R_S^2}{R^2}\right)\,\epsilon_{n,m}^D\,,
\label{eq:eccgranular-app}
\end{align}
for all $n \neq 0$.

So we can expect that $m=4$ eccentricities, such as $\epsilon_{2,4}$,  are approximately twice as sensitive to the size of the source as $m=2$ eccentricities, such as the traditional $\epsilon_2 = \epsilon_{2,2}$ that determines the leading elliptic flow.
Further, with this formula we can quantify what length scales can be probed.  For example, if the granular structure were different by a fraction of the system size $R_S/R = X \ll 1$,  the fractional change in $\epsilon_{2,4}$  would be roughly $2X^2$.  
For $Pb+Pb$ collisions at $0-10\%$ centrality, for instance, where the subleading $v_2$ mode is dominated by $\epsilon_{2,4}$ our \trento initial conditions yield  $R_D\sim 4$ fm. 
So if we want to probe the system at scales smaller than, e.g., 1 fm, Eq.~\eqref{eq:eccgranular-app} suggests 
we need to measure the subleading $v_2$ PCA mode to no better than $\sim 10\%$ precision. 
 An even larger effect is to be expected for smaller collision systems and for higher harmonics, where higher values of $m$ become relevant.

To verify these expectations from the simple independent source model, we simulate  realistic \trento events, but varying the Gaussian width $w$ of the nucleon, which sets the granularity scale in the model \cite{Moreland:2014oya}.
For each initial condition, we calculated the root-mean-square radius 
$R = \sqrt{\{ r^2\} }= \sqrt{W_{0,2}}$ and the eccentricities  
$\epsilon_{2,2}$ and  $\epsilon_{2,4}$ for $w = 0.4 - 0.8$ fm and $w = 0.956$ fm 
--- the latter corresponding to the optimum value found in Ref.~\cite{Bernhard:2018hnz}. 
For each value of $w$, these initial conditions were ordered according to their total entropy content 
and binned into 10 quantiles, playing the role of centrality bins.

Figure~\ref{fig:Eps24Sigma} shows the leading and subleading 
elliptical eccentricities of the generated initial conditions as functions of 
$w$. 
Because principal component analysis isolates linearly uncorrelated fluctuations, 
we define 
\begin{equation}
 \hat\epsilon_{2,4} \equiv \epsilon_{2,4} - \dfrac{\langle \epsilon_{2,2}^* \,\epsilon_{2,4} \rangle}{\langle|\epsilon_{2,2}|^2\rangle}\,\epsilon_{2,2}\,,
\end{equation}
and show 
\begin{equation}
 \epsilon_{2,2}\{2\}\equiv \sqrt{\langle|\epsilon_{2,2}|^2\rangle}\,,\textrm{ and }\, 
  \hat\epsilon_{2,4}\{2\}\equiv \sqrt{\langle|\hat\epsilon_{2,4}|^2\rangle}
\end{equation}
for three different centralities, between $0\%$ and $50\%$. 
Under a decrease of $w$ from $0.956$ fm to $0.4$ fm, $\sqrt{\langle |\hat\epsilon_{2,4}|^2\rangle}$ was found to increase by roughly $35-80\%$, 
depending on centrality, with $\sqrt{\langle |\epsilon_{2,2}|^2\rangle}$ changing by about $14-22\%$. 
Even in central collisions, where sensitivity to the nucleon-width $w$ was found to be smaller, 
$\sqrt{\langle|\hat\epsilon_{2,4}|^2\rangle}$ is more than twice as sensitive to $w$ as 
 $\sqrt{\langle|\epsilon_{2,2}|^2\rangle}$, surpassing expectations from Eq.~\eqref{eq:eccgranular-app}.  

To compare \trento results  to expectations from the isotropic ISM,  we assume $R_S = \sqrt{\{x^2 + y^2\}}_S \approx \sqrt{2}\,w$ in Eq.~\eqref{eq:eccgranular-ex}. 
Because the root-mean square radius $R$ is found to be mildly dependent on $w$, we assume the
  source-distribution r.m.s. radius, $R_D = \sqrt{R^2 - R_S^2}$, to be held constant, so that
\begin{equation}
   \dfrac{\epsilon_{n,m}(w')}{\epsilon_{n,m}(w)} \stackrel{ISM}{\approx} \left(\dfrac{R^2(w)}{R^2(w) + 2\,(w'^2-w^2)}\right)^{m/2}\,\epsilon_{n,m}^D\,.
   \label{eq:eccgranular-Rd}
\end{equation}
In Fig.~\ref{fig:Eps24Sigma}, Eq.~\eqref{eq:eccgranular-Rd} is represented, for $40-50\%$ centrality, by the faint magenta lines. 
A reasonable agreement is found between the isotropic ISM and \trento  for small departures from $w=0.956$, but significant deviations 
are seen for $w$ below $0.6$ fm. 
This might be explained by anisotropies at the granular scale, which should be expected from the 
generalized average involved in the \trento reduced thickness function --- a combination of projectile and target thicknesses which is 
nonlinear for parameter $p\neq 1$  \cite{Moreland:2014oya}. 
Nonetheless, we note that a much better agreement with the isotropic ISM can be found by employing $\epsilon_{2,4}$ --- with no subtraction of correlations with $\epsilon_{2,2}$ --- instead of 
$\hat \epsilon_{2,4}$. 
We also stress that even the results for Eq.~\eqref{eq:eccgranular-Rd} exceed the twofold increase in sensitivity from $m=2$ to $m=4$ expected from 
Eq.~\eqref{eq:eccgranular-app}, because of departures from linear behavior. 

Above, we have presented analytical and numerical evidence for the higher sensitivity of the higher-order eccentricity $\epsilon_{2,4}$ to the granularity of initial-state fluctuations. Even under the very conservative assumption of isotropic fluctuations 
at small scales, this eccentricity is found to be at least two times more sensitive 
to the granular scale $w$ than the usual eccentricity $\epsilon_{2,2}$.   
We thus conclude that the subleading principal component of elliptic flow $V_2^{(2)}(p_T)$, being strongly sensitive to $\sqrt{\langle|\hat\epsilon_{2,4}|^2\rangle}$,  
provides a unique probe of the initial conditions at sub-Fermi scales.

\section{Conclusions}
\label{sec:conclusions}

In this paper, we show that the PCA of anisotropic flow is a promising tool for studying the hydrodynamic response of the QGP to fluctuations of the initial state. 
Once undesired contributions from radial flow fluctuations are properly removed, as proposed in \cite{Hippert:2019swu}, this analysis uncovers   
details of the hydrodynamic response to anisotropies of the initial energy distribution of the system. 
More specifically, it reveals corrections to the familiar scaling relation between flow harmonics and spatial 
eccentricities $V_n = \kappa_n\, \epsilon_n$ \cite{Teaney:2010vd,Gardim:2011xv,Teaney:2012ke,Gardim:2014tya,Fu:2015wba,Rao:2019vgy,Qin:2010pf,Qiu:2011iv,Niemi:2012aj,Giacalone:2017uqx,Wei:2018xpm}. 

A more complete mapping between initial geometry and flow harmonics can be generalized by means of a cumulant expansion of the initial 
transverse energy density profile \cite{Teaney:2010vd,Gardim:2011xv,Teaney:2012ke,Gardim:2014tya,Fu:2015wba,Rao:2019vgy}. In this work, by extending this mapping to account for transverse-momentum dependence, 
we were able to successfully predict anisotropic flow harmonics on a differential basis. 
This allowed us to systematically gauge, for the first time, the relative importance of linear and 
nonlinear hydrodynamic response at different centralities and transverse-momentum ranges. 
Higher-order cumulants of the initial transverse profile of the system proved to be more important 
at lower transverse momentum and in more central collisions, while 
nonlinear hydrodynamic response was found to provide the most important corrections 
at higher transverse momentum and in more peripheral collisions. 
We also found that two nonlinear terms, proportional to $|\epsilon_{2,2}|^2\epsilon_{2,2}$ 
and $\epsilon_{1,3}^2$, provide relevant corrections to $V_2(p_T)$. 
Surprisingly, the latter was found to surpass the former in importance 
on a wide transverse-momentum range.

By predicting $V_n(p_T)$ exclusively from features of the initial transverse geometry, we were also able to reproduce,  
to a reasonable accuracy, the principal components of elliptic and triangular flow calculated in full event-by-event 
hydrodynamic simulations. 
By employing different eccentricities of the initial geometry, we found subleading principal components to be sensitive both to 
higher-order cumulants of the initial transverse geometry and to nonlinear hydrodynamic response. 
In the case of triangular flow, the most important contribution to the subleading component comes from 
a nonlinear term proportional to $\epsilon_{2,2}\epsilon_{1,3}$. 
The leading linear correction, proportional to $\epsilon_{3,5}$, on the other hand, 
was found to provide distinct contributions to the first subleading component
--- qualitatively different from the full simulation results. 

In the case of elliptic flow fluctuations, the leading linear correction to $V_2(p_T)$, proportional 
to $\epsilon_{2,4}$, was found to provide an excellent prediction of the first subleading principal component 
in central collisions. In more peripheral collisions, this term was found to compete with 
the nonlinear terms $\propto|\epsilon_{2,2}|^2\epsilon_{2,2}$ and $\propto\epsilon_{1,3}^2$ in importance. 
By employing both analytical arguments and numerical results, 
we showed this eccentricity to be sensitive to the initial granularity of the system. 
This provides compelling evidence that the measurement of the PCA observables proposed in Ref.~\cite{Hippert:2019swu} 
will shed light on the details and granular structure of the initial stages of high-energy 
nucleus-nucleus collisions.

In short, measurements of the principal components of different flow harmonics at different centralities 
can be employed to study different aspects of the hydrodynamic response of the QGP. 
In addition to fluctuations of the energy density at smaller scales, principal component analysis can prove a useful tool 
to uncover fluctuations originating from initial flow \cite{Sousa:2020cwo}. 
This makes PCA an especially interesting tool for investigating the physics of 
smaller collision systems \cite{Schenke:2019pmk} --- although a study of this possibility is left to future work.

\section*{Acknowledgments}

We are thankful to G.~S.~Denicol, D.~Teaney, and A.~Mazeliauskas  for fruitful discussions. 
This research was funded by FAPESP Grants No. 2016/13803-2 (D.D.C.),
No. 2016/24029-6 (M.L.), No. 2017/05685-2 (all), No. 2018/01245-0 (T.N.dS.), No. 2018/07833-1 (M.H.), and  No. 2019/16293-3 (J.G.P.B.). 
D.D.C., M.L., and J.T. thank CNPq for financial support. 
J.N. is partially supported by the U.S. Department of Energy, Office of Science, Office for Nuclear Physics under Award No. DE-SC0021301. 
This research used the computing resources and assistance of the John David
Rogers Computing Center (CCJDR) in the Institute of Physics "Gleb
Wataghin," University of Campinas.
  \bibliography{mappingPCA}

\appendix

 \section{Moments and Cumulants of the Initial Density Profile}
 \label{app:cumsandmoms}
 
 In this appendix, we provide a complementary, more detailed description of the cumulants and moments 
 of the initial transverse energy-density profile, used to define the eccentricities in Eq.~\eqref{eq:defecc}. 
 
 \subsection{Definition}
 
 The two-dimensional density (here, energy-density) profile of the initial state $\rho(\vec x)$ can be characterized 
by its moments and cumulants. 
The corresponding generating functions are, respectively, 
\begin{equation}
 \rho(\vec k) = \int d^2 x\, \rho(\vec x)\, e^{i\vec k \cdot \vec x}
\end{equation}
and 
\begin{equation}
 W(\vec k) = \log \big(\rho(\vec k)/\bar\rho\big)\,,
  \label{eq:genfunc}
\end{equation}
where $\bar\rho$ is an arbitrary scale and $\vec x = r\,(\cos \phi_x, \sin \phi_x)$ and $\vec k = k\,(\cos \phi_k, \sin \phi_k)$ 
are vectors on the transverse plane. 
The moments and cumulants can be defined by taking the Taylor expansion of both $\rho(\vec k)$ and $W(\vec k)$ around $k\equiv|\vec k|=0$, followed by a 
Fourier expansion in $\phi_k$:
\begin{align}
 \rho(\vec k ) &= \sum_{m=0}^\infty \sum_{n=-m}^{m} \varrho_{n,m}\,k^m\,e^{-i n \phi_k}\,,\\
  W(\vec k ) &= \sum_{m=0}^\infty \sum_{n=-m}^{m} \mathcal{W}_{n,m}\,k^m\,e^{-i n \phi_k}\,,
 \label{eq:genfunc2}
 \end{align}
 where, from Eqs.~\eqref{eq:seriesrho} and \eqref{eq:seriesW}, we can identify
 \begin{equation}
\varrho_{n,m}\equiv \dfrac{(i/2)^m\,\rho_0\, \rho_{n,m}}{\left(\frac{m+n}{2}\right)!\,\left(\frac{m-n}{2}\right)!}\,, 
\qquad
\mathcal{W}_{n,m}\equiv \dfrac{(i/2)^m\, W_{n,m}}{\left(\frac{m+n}{2}\right)!\,\left(\frac{m-n}{2}\right)!}\,.
\label{eq:newdefrhoW}
 \end{equation}

Note that
\begin{align}
\begin{split}
  \varrho_{n,m} &= \frac{1}{m!}\,\int_{-\pi}^{\pi} \frac{d\phi_k}{2\pi}\,\frac{d^m\;}{dk^m} \rho(\vec k) e^{in\phi_k}\bigg|_{\vec k =0} \\
 &= \frac{i^m}{m!}\,\int d^2 x\,\rho(\vec x)\,r^m\, \int_{-\pi}^{\pi} \frac{d\phi_k}{2\pi}\,[\cos(\phi_k-\phi_x)]^m\,e^{in\phi_k}\,.
\end{split}
 \end{align}
Using the result,
\begin{align}
\begin{split}
I(\phi_x) &\equiv  \int \frac{d\phi_k}{2\pi}\,[\cos(\phi_k-\phi_x)]^m\,e^{in\phi_k} \\
&= \frac{1}{2^m} \,e^{in\phi_x}\,\int_{-\pi}^{\pi} \frac{d\phi}{2\pi}\,(e^{i\phi}+e^{-i\phi})^m\,e^{in\phi}\\
 &= \frac{1}{2^m} \,e^{in\phi_x}\sum_{l=0}^m \frac{m!}{l!(m-l)!}\,\int_{-\pi}^{\pi} \frac{d\phi}{2\pi}\,e^{-i(2l-m)\phi}\,e^{in\phi}\\
 &= \frac{m!}{2^m\,(\frac{m+n}{2})!(\frac{m-n}{2})!} \,e^{in\phi_x}\,, 
\label{eq:rhoIntegral}
\end{split}
\end{align}
we get 
\begin{align}
 \begin{split}
  \rho_{n,m}&= \frac{1}{\rho_0} \,\int d^2 x\,\rho(\vec x)\,r^m\,e^{in\phi_x}\\
 &= \left\{ r^m\,e^{in\phi_x} \right\}
  \,,
    \label{eq:momentsredef} 
 \end{split}
\end{align}
provided $m-|n|$ is a positive multiple of $2$. 
For negative or odd values of $m-|n|$, the integral in Eq.~\eqref{eq:rhoIntegral} vanishes.

 \subsection{Relation between cumulants and moments}

 The cumulants can be written in terms of the moments by separating the zero mode 
 and using the Taylor expansion of $\log h$ around $h=1$:
\begin{align}
\begin{split}
 W(\vec k) - W(\vec0)
  ={}& \sum_{l=1}^\infty \frac{(-1)^{\ell-1}}{\ell} \left( \sum_{m=1}^\infty \sum_{n=-m}^{m} \frac{\varrho_{n,m}}{\rho_0}\,k^m\,e^{-i n \phi_k}\right)^\ell \\
  ={}&  \sum_{l=1}^\infty \frac{(-1)^{\ell-1}}{\ell\,(\rho_0)^\ell} \sum_{\sum p_{n,m}=\ell } \ell!\times\\
  &\times \prod_{m=0}^\infty\prod_{n=-m}^m \frac{1}{p_{n,m}!}\left({\varrho_{m,n}} k^m \,e^{-i n \phi_k}\right)^{p_{n,m}}\\
  = {}& \sum_{\ell=1}^\infty \frac{(-1)^{\ell-1} (\ell-1)!}{(\rho_0)^\ell}\times \\
  &\times \sum_{\substack{\{p_{n,m}\}\\\sum p_{n,m}=\ell }} \,
  k^{\,\sum m\,p_{n,m}}\,e^{\,i\sum n\,p_{n,m}\,\phi_k}\times\\ 
  &\times \prod_{m=0}^\infty\prod_{n=-m}^m \frac{1}{p_{n,m}!}({\varrho_{m,n}} )^{p_{n,m}}\,,
\label{eq:multnomexpW}
  \end{split}
\end{align}
where the multinomial theorem was employed to rewrite the $\ell$th power of a sum as a sum over the powers $p_{n,m}$ of each term, under the condition that 
$\sum_{m,n} p_{n,m}=\ell$. 

Equating the powers of $k$ and $e^{i\phi_k}$ in Eqs.~(\ref{eq:genfunc2}) and (\ref{eq:multnomexpW}) allows us to identify the cumulants,
\begin{align}
\begin{split}
 \mathcal{W}_{\bar n,\bar m} ={}& \sum_{\substack{\{ p_{n,m} \}\\ \substack{\sum m\,p_{n,m} = \bar m}\\\sum n\,p_{n,m} = \bar n}}
                     (-1)^{\left(\sum_{n,m} p_{n,m}-1\right)} \times \\
                     &\;\;\; \times \left( \sum_{m=0}^\infty\sum_{n=-m}^m   p_{n,m}-1 \right)!\times \\
                     &\times \prod_{m=0}^\infty\prod_{n=-m}^m  \frac{1}{p_{n,m}!}\left(\frac{\varrho_{m,n}}{\rho_0} \right)^{p_{n,m}}         \,,
\label{eq:cumulant1}
\end{split}
\end{align}
for $\bar m\neq 0$,
while, for $\bar m = 0$, we find simply 
\begin{equation}
 \mathcal{W}_{0,0} = \log (\rho_0/\bar\rho)\,.
\end{equation}

Equation~\eqref{eq:cumulant1} gives us the cumulants $\mathcal{W}_{\bar n,\bar m}$ in terms of the moments $\varrho_{m,n}$. 
The sum is over all possible partitions  of $r^{\bar m} \,e^{i\bar n\phi_x}$, with weights given by 
\begin{equation}
\tilde\Omega^{\bar m \bar n}_{m,n}= (-1)^{\left(\sum_{n,m} p_{n,m}-1\right)} \,\left( \sum_{m,n}   p_{n,m}-1 \right)!\,\prod_{n,m} \left(p_{n,m}!\right)^{-1}\,.
\label{eq:weightmod}
\end{equation}
Using Eqs.~\eqref{eq:newdefrhoW} and \eqref{eq:momentsredef}, we can finally find ${W}_{\bar n,\bar m}$ in terms of $\rho_{m,n}$:
\begin{widetext}
  \begin{equation}
 W_{\bar n,\bar m \neq 0} = \left(\tfrac{\bar m+ \bar n}{2}\right)!\,\left(\tfrac{\bar m-\bar n}{2}\right)!\, \sum_{\substack{\{ p_{n,m} \}\\ \substack{\sum m\,p_{n,m} = \bar m}\\\sum n\,p_{n,m} = \bar n}}
                     (-1)^{\left(\sum_{n,m} p_{n,m}-1\right)} \, \left( \sum_{m=0}^\infty\sum_{n=-m}^m   p_{n,m}-1 \right)!
                     \, \prod_{m=0}^\infty\prod_{n=-m}^m  \frac{\left\{ r^m\,e^{in\phi_x} \right\}^{p_{n,m}} }{p_{n,m}![(\frac{m+n}{2})!(\frac{m-n}{2})!]^{p_{n,m}}}        \,,
\label{eq:finalcml}
\end{equation}
\end{widetext}
where only even values of $m-|n|$ are considered. 
Once again, the sum is over all possible partitions, with $\sum m\,p_{n,m} = \bar m$ and $\sum n\,p_{n,m} = \bar n$. 
The weight of each term is of the form,
\begin{equation}
\Omega^{\bar m \bar n}_{m,n}= \tilde\Omega^{\bar m \bar n}_{m,n}\times \dfrac{\left(\tfrac{\bar m+ \bar n}{2}\right)!\,\left(\tfrac{\bar m-\bar n}{2}\right)!}{[(\frac{m+n}{2})!(\frac{m-n}{2})!]^{p_{n,m}}}\,.
\label{eq:weight}
\end{equation}
From Eq.~\eqref{eq:finalcml} --- or, equivalently, from Eqs.~\eqref{eq:weightmod} and \eqref{eq:weight} and all the relevant partitions --- 
one can recover Eqs.~\eqref{eq:W02}, \eqref{eq:W22}, \eqref{eq:W13} and \eqref{eq:W33}. 

\section{Response Coefficients}
\label{app:coeffs}

In Sec.~\ref{sub:mappresponse}, we have described how the response coefficients $\kappa^{(n)}_{\{m',n'\}}$ were extracted for different estimators of the flow harmonics. Predictions of the flow harmonics and their principal components from these estimators were shown in Sec.~\ref{sec:results}, but, for simplicity, the response coefficients themselves were not shown as functions of $p_T$. 
Here, we show the centrality and transverse-momentum dependence of the coefficients of the leading and subleading terms, for different predictors of elliptic and triangular flow. 

\begin{figure*}
 \centering
 \includegraphics[width=\textwidth]{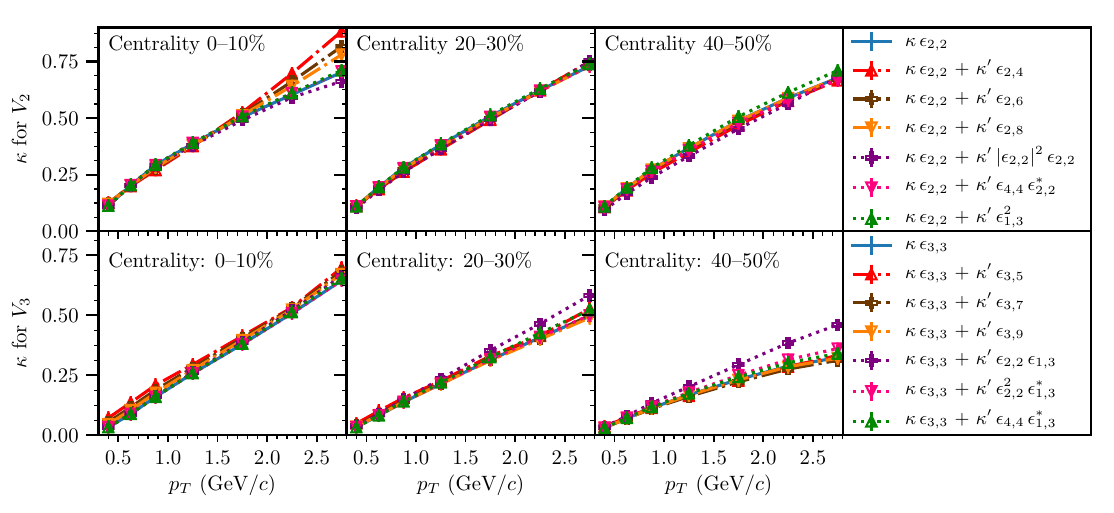}
 \caption{ Response coefficient $\kappa(p_T)$ for the leading terms $\kappa\,\epsilon_{2,2}$ and $\kappa\,\epsilon_{3,3}$, corresponding to  elliptic and triangular flow, respectively. Different curves correspond to different predictors, but the leading eccentricity is kept as $\epsilon_{n,n}$. 
 Events are simulated for $Pb+Pb$ collisions at $\sqrt{s_{NN}}=2.76$ TeV, within a hybrid event-by-event hydrodynamic model (\trentonosp +\textsc{Music}+UrQMD).  }
 \label{fig:kappalead23}
\end{figure*}

Figure~\ref{fig:kappalead23} shows the response coefficient $\kappa(p_T)$ for the leading term, $\kappa(p_T)\,\epsilon_{n,n}$, in these predictors. This response coefficient is responsible for the shape of the leading principal component $V_n^{(1)}(p_T)$ of the flow harmonics. In general, the leading and subleading terms in the predictors can be correlated, so that different choices of subleading term can slightly affect  $\kappa(p_T)$. 

\begin{figure*}
 \centering
 \includegraphics[width=\textwidth]{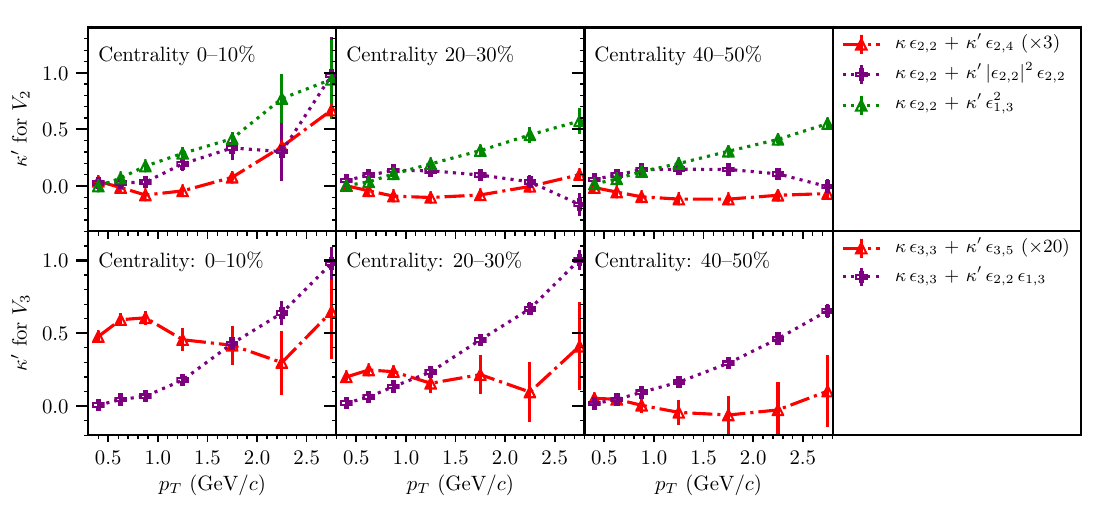}
 \caption{ Response coefficient $\kappa'(p_T)$ for the subleading term in the predictors for  elliptic and triangular flow. Different curves correspond to different predictors. 
 Events are simulated for $Pb+Pb$ collisions at $\sqrt{s_{NN}}=2.76$ TeV, within a hybrid event-by-event hydrodynamic model (\trentonosp +\textsc{Music}+UrQMD).  }
 \label{fig:kappasub23}
\end{figure*}

Figure~\ref{fig:kappasub23} shows the response coefficient $\kappa'$ for different choices of subleading term. Different terms correlate better with different transverse-momentum regions, yielding the different shapes in $\kappa'(p_T)$. The shape of this response coefficient is especially important for the shape of the subleading principal component $V_n^{(2)}(p_T)$ of the flow harmonics.

\section{Other Harmonics and Corrections}
\label{app:otherharms}

\begin{figure*}
 \centering
 \includegraphics[width=\textwidth]{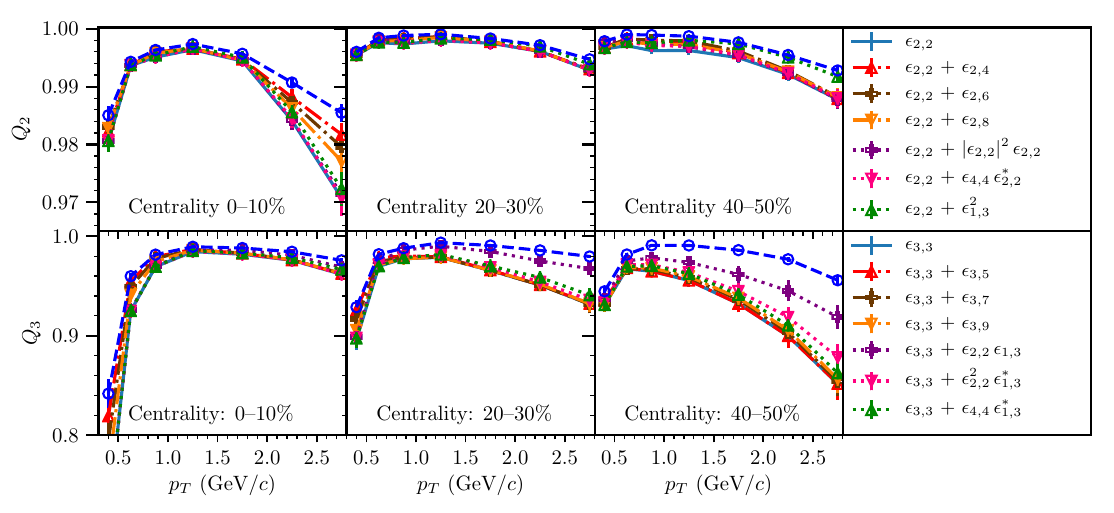}
 \caption{ Pearson correlation coefficient between the flow harmonics $V_2(p_T)$ (upper panel) and $V_3(p_T)$ (lower panel) and predictions of their event-by-event fluctuations from eccentricities of the 
 initial transverse geometry. Different curves correspond to different predictors, with the blue dashed curve corresponding to the full expressions in Eqs.~\eqref{eq:predV2} and \eqref{eq:predV3}. 
 Events are simulated for $Pb+Pb$ collisions at $\sqrt{s_{NN}}=2.76$ TeV, within a hybrid event-by-event  hydrodynamic model (\trentonosp +\textsc{Music}+UrQMD). }
 \label{fig:Pearson23_ext}
\end{figure*}

For the sake of clarity and simplicity, we have opted to omit a few results from the main text of this article. 
This had the advantage of making the text paper clearer and the plots less polluted. 
For completeness, we present a few extra results in this appendix. 

Here, we show more comprehensive results for the mapping of the anisotropic flow from eccentricities of the initial geometry. 
In all these results,  
the dashed blue curve with hollow circles represents predictions from the most complete estimates available. 
For $n=2,3$, these estimates can be found on Eqs.~\eqref{eq:predV2} and \eqref{eq:predV3}. 
For $n=4,5$, they are given by
\begin{align}
\begin{split}
    V_4 \simeq {}& \kappa^{(4)}_4\,\epsilon_{4,4}  
 +  \kappa^{(4)}_6\,\epsilon_{4,6} +\kappa^{(4)}_8\,\epsilon_{4,8} + \mathcal{O}(m=10)\\
 &+    \kappa^{(4)}_{(2,2)^2}\,\epsilon_{2,2}^2 +   \kappa^{(4)}_{\substack{(1,3)\\(3,3)}}\,\epsilon_{1,3}\,\epsilon_{3,3} \\
  &+ \ldots  
  + \mathcal{O}(\epsilon^3)\,,
  \label{eq:predV4}
\end{split}
\end{align} 
\begin{align}
 \begin{split}
  V_5 \simeq {}& \kappa^{(5)}_5\,\epsilon_{5,5} 
  +  \kappa^{(5)}_7\,\epsilon_{5,7} +\kappa^{(5)}_9\,\epsilon_{5,9} + \mathcal{O}(m=11)\\
  &+    \kappa^{(5)}_{\substack{(2,2)\\(3,3)}}\,\epsilon_{2,2}\,\epsilon_{3,3}
  + \kappa^{(5)}_{\substack{(1,3)\\(4,4)}}\,\epsilon_{1,3}\,\epsilon_{4,4}\\  &+ \ldots
  + \mathcal{O}(\epsilon^3)\,.
 \label{eq:predV5}    
 \end{split}
\end{align}
Wherever present, the solid black curve with hollow squares represents the results from full hydrodynamic simulations. 
Dot-dashed curves represent results from predictors combining the leading eccentricity $\epsilon_{n,n}$ 
and a single linear correction $\propto \epsilon_{n,m>n}$. 
Dotted curves, on the other hand, exhibit results from predictors combining the leading eccentricity $\epsilon_{n,n}$ 
and a single nonlinear subleading term. 

\begin{figure*}
 \centering
 \includegraphics[width=\textwidth]{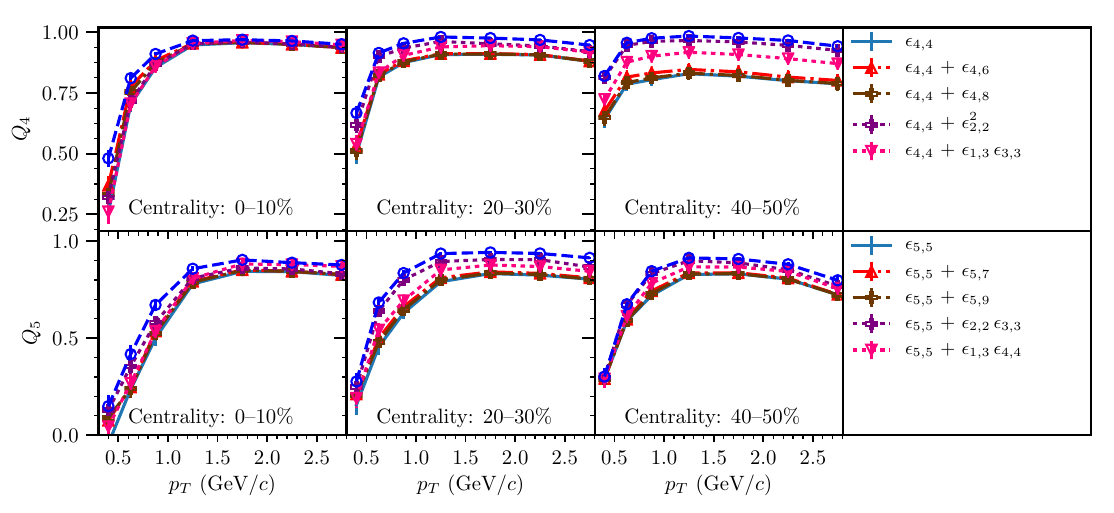}
 \caption{ Pearson correlation coefficient between the flow harmonics $V_4(p_T)$ (upper panel) and $V_5(p_T)$ (lower panel) and predictions of their event-by-event fluctuations from eccentricities of the 
 initial transverse geometry. Different curves correspond to different predictors, with the blue dashed curve corresponding to the full expressions in Eqs.~\eqref{eq:predV2} and \eqref{eq:predV3}. 
 Events are simulated for $Pb+Pb$ collisions at $\sqrt{s_{NN}}=2.76$ TeV, within a hybrid event-by-event hydrodynamic model (\trentonosp +\textsc{Music}+UrQMD). }
 \label{fig:Pearson45_ext}
\end{figure*}

In Sec.~\ref{sec:predQ} we have shown results for the quality of different 
predictors of the elliptic and triangular flow harmonics. 
Figure~\ref{fig:Pearson23}, in particular, displays the Pearson correlation coefficient between  the actual flow harmonics 
from hydrodynamic simulation events and predictors of these harmonics from the initial-state eccentricities. 
In Fig.~\ref{fig:Pearson23_ext}, we repeat the same results, but, for completeness include other predictors 
of $V_2(p_T)$ and $V_3(p_T)$, each of them built from a pair of terms containing the leading term $\propto\epsilon_{n,n}$. 
Figure~\ref{fig:Pearson45_ext} exhibits the same kind of analysis for $V_4(p_T)$ and $V_5(p_T)$. 
It is noteworthy that our predictions for higher harmonics are not quite as good as the ones for elliptic and triangular flow.  Note also that the linear estimator is better here than in some previous analyses because of our choice to define eccentricities via cumulants rather than moments.

\begin{figure*}
 \centering
 \includegraphics[width=\textwidth]{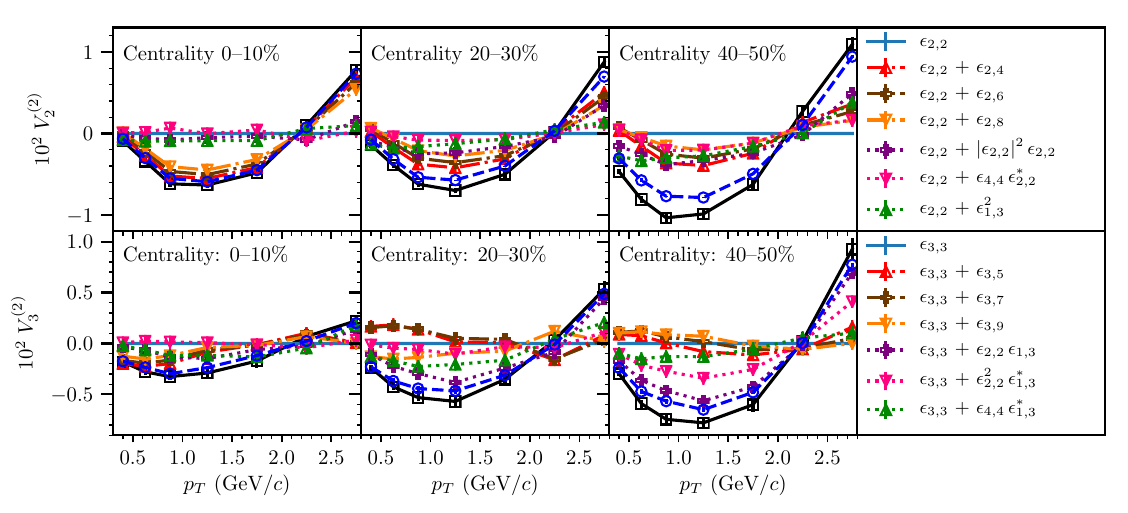}
 \caption{ Second principal component of elliptic (upper panel) and triangular (lower panel) flow, both from full hydrodynamic simulations and event-by-event predictions from eccentricities of the 
 initial geometry. Different curves correspond to different predictors, with the black solid curve corresponding to the full hydrodynamic results and the blue dashed curve corresponding to the full 
 expressions in Eqs.~\eqref{eq:predV2} and \eqref{eq:predV3}. 
 Events are simulated for $Pb+Pb$ collisions at $\sqrt{s_{NN}}=2.76$ TeV, within a hybrid event-by-event hydrodynamic model (\trentonosp +\textsc{Music}+UrQMD).  }
 \label{fig:PCAmapp23_extsub}
\end{figure*}

\begin{figure*}
 \centering
 \includegraphics[width=\textwidth]{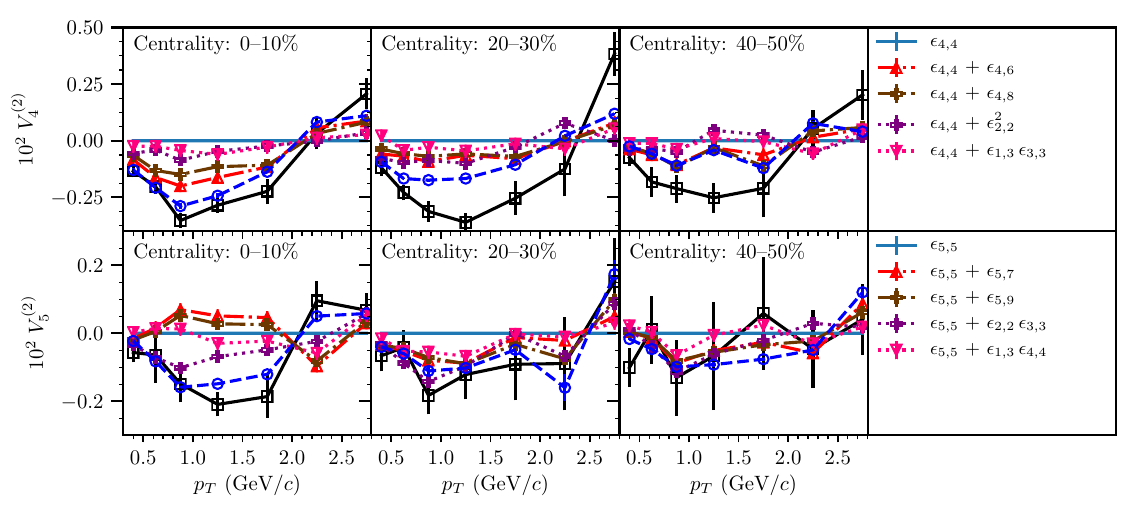}
 \caption{ Second principal component of $V_4(p_T)$ (upper panel) and $V_5(p_T)$ (lower panel), both from full hydrodynamic simulations and event-by-event predictions from eccentricities of the 
 initial geometry. Different curves correspond to different predictors, with the black solid curve corresponding to the full hydrodynamic results and the blue dashed curve corresponding to the full 
 expressions in Eqs.~\eqref{eq:predV4} and \eqref{eq:predV5}. 
 Events are simulated for $Pb+Pb$ collisions at $\sqrt{s_{NN}}=2.76$ TeV, within a hybrid event-by-event hydrodynamic model (\textsc{TrENTo}+\textsc{Music}+UrQMD).  }
 \label{fig:PCAmapp45_extsub}
\end{figure*}

In Sec.~\ref{sec:mappPCA} and, more specifically, in Fig.~\ref{fig:PCAmapp23}, we have presented results for the second principal component 
of elliptic and triangular flow harmonics. 
There, we plotted results from full event-by-event hydrodynamic simulations, from the predictors in Eqs.~\eqref{eq:predV2} and \eqref{eq:predV3}, 
and from a few predictors containing a pair of terms each.
In Fig.~\ref{fig:PCAmapp23_extsub}, we, once again, plot the first subleading principal component of elliptic and triangular flow. 
However, in this figure, we include other predictors as well. 
In Fig.~\ref{fig:PCAmapp45_extsub}, we show the first subleading principal component of $V_4(p_T)$ and $V_5(p_T)$. 
For these higher harmonics, results are less impressive as higher harmonics likely require  a larger number of terms to describe.

\begin{figure*}
 \centering
 \includegraphics[width=\textwidth]{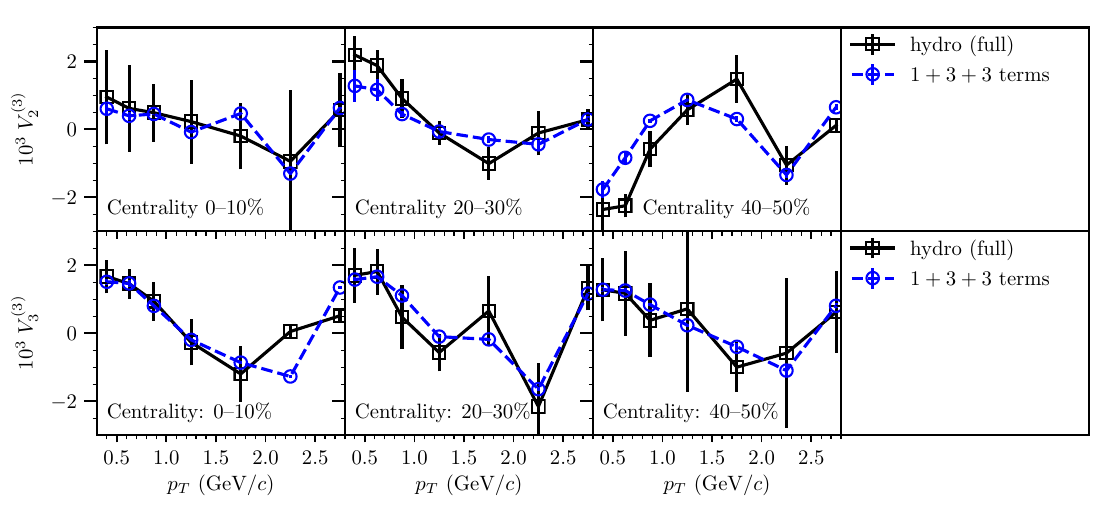}
 \caption{ Third principal component of elliptic (upper panel) and triangular (lower panel) flow, both from full hydrodynamic simulations and event-by-event predictions from eccentricities of the 
 initial geometry. The black solid curve corresponds to the full hydrodynamic results, while the blue dashed curve corresponds to the full 
 expressions in Eqs.~\eqref{eq:predV2} and \eqref{eq:predV3}. 
 Events are simulated for $Pb+Pb$ collisions at $\sqrt{s_{NN}}=2.76$ TeV, within a hybrid event-by-event hydrodynamic model (\trentonosp +\textsc{Music}+UrQMD).  }
 \label{fig:PCAmapp23subsub}
\end{figure*}

Finally, we have also calculated principal components beyond the first subleading one. 
Figure~\ref{fig:PCAmapp23subsub} shows the third principal components of triangular and elliptic flow fluctuations. 
Results are shown only for full hydrodynamic simulations and for the full predictors on Eqs.~\eqref{eq:predV2} and \eqref{eq:predV3}.  
The agreement between the two curves is quite striking, considering the level of detail 
captured by the third principal component.

\nocite{Tange2011a}

\end{document}